\documentclass[aps,prd,twocolumn,groupedaddress,nofootinbib]{revtex4-1}

\usepackage{graphicx,color,amsmath}

\usepackage{dcolumn}
\usepackage{multirow}
\usepackage{tabularx}
\usepackage{verbatim}
\usepackage{tablefootnote}
\usepackage{hyperref}

\newcommand{\second}{{\, {\rm s}}}

\newcommand{\um}{{\, {\rm \mu m}}}

\newcommand{\cm}{{\, {\rm cm}}}

\newcommand{\mm}{{\, {\rm mm}}}
\newcommand{\mum}{{\, \mu{\rm m}}}
\newcommand{\micron}{{\, \mu{\rm m}}}

\newcommand{\meV}{{\, {\rm meV}}}
\newcommand{\eV}{{\, {\rm eV}}}
\newcommand{\keV}{{\, {\rm keV}}}

\newcommand{\GeV}{{\, {\rm GeV}}}

\newcommand{\kelvin}{{\, {\rm K}}}
\newcommand{\K}{{\, {\rm K}}}

\newcommand{\Hz}{{\, {\rm Hz}}}

\newcommand{\watt}{{\, {\rm W}}}

\newcommand{\LL}{{\mathcal{L}}}

\newcommand{\OO}{{\mathcal{O}}}

\definecolor{mypurple}{RGB}{164,64,214}
\newcommand{\rnlcol}[1]{}

\begin{document}

\title{Axion and hidden photon dark matter detection with multilayer optical haloscopes}

\author{Masha Baryakhtar}
\email{mbaryakhtar@perimeterinstitute.ca}
\affiliation{Perimeter Institute for Theoretical Physics, 31 Caroline Street North, Waterloo, Ontario N2L 2Y5, Canada}
\author{Junwu Huang} 
\email{jhuang@perimeterinstitute.ca}
\affiliation{Perimeter Institute for Theoretical Physics, 31 Caroline Street North, Waterloo, Ontario N2L 2Y5, Canada}
\author{Robert Lasenby} 
\email{rlasenby@perimeterinstitute.ca}
\affiliation{Perimeter Institute for Theoretical Physics, 31 Caroline Street North, Waterloo, Ontario N2L 2Y5, Canada}

\date{\today}

\begin{abstract}

 A well-motivated class of dark matter candidates, including axions
 and dark photons, takes the form of coherent oscillations of
 a light bosonic field. If the dark matter couples to Standard Model
 states, it may be possible to detect it via absorptions in a laboratory
 target. Current experiments of this kind include cavity-based
 resonators that convert bosonic dark matter to electromagnetic fields,
 operating at microwave frequencies. We propose a new class of detectors
 at higher frequencies, from the infrared through the ultraviolet, based
 on the dielectric haloscope concept. In periodic photonic materials,
 bosonic dark matter can efficiently convert to detectable single
 photons. With feasible experimental techniques, these detectors can
 probe significant new parameter space for axion and dark
 photon dark matter in the $0.1-10\eV$ mass range.
\end{abstract}

\maketitle


\section{Introduction}

There is overwhelming evidence that the majority of the
matter density of the universe takes some beyond-Standard-Model
form, referred to as dark matter (DM)~\cite{Rubin:1970zza,Spergel:2003cb}.
Despite this, the form of the dark matter remains almost entirely
unknown. If, like Standard Model (SM) matter, it is
a relic from the hot thermal plasma of the early universe,
then the fact that it is `cold' (low-velocity) today means
that it cannot consist of particles lighter than $\sim \keV$~\cite{Yeche:2017upn,Irsic:2017ixq}.
However, there are a range of alternative, non-thermal production
mechanisms which could generate a viable cold dark population
of lighter particles. In many beyond-the-Standard-Model theories, these light
new particles can naturally have very small couplings to SM
states, allowing them to be stable and hard-to-detect. Thus, such particles can serve
as attractive dark matter candidates.
At masses $\lesssim 100 \eV$, the dark matter must be bosonic,
since the Pauli exclusion principle forbids fermionic DM
from having the dense, low-velocity distributions observed
in galaxies~\cite{Boyarsky:2008ju}.

Unless there is a symmetry preventing it, the leading-order
interaction between light bosonic dark matter and Standard Model matter will be
 absorption and emission of single DM particles. This is true for the simplest and most attractive
light DM models, such as axions~\cite{axion1,axion2,axion3} or dark photons~\cite{Holdom:1985ag}.
Accordingly, a range of existing and proposed experiments aims
to detect the absorption of light DM through different mechanisms
(see~\cite{Irastorza:2018dyq,Graham:2015ouw,Jaeckel:2010ni} for reviews of axion and dark photon DM detection experiments).
However, many of these are not sensitive to DM masses far above
the microwave frequency range. In this paper, we discuss how
to extend the search for light dark matter candidates
to higher masses, from $0.1$ to $10 \eV$. 

For many kinds of dark matter couplings, DM to photon conversion
is a promising experimental approach, transferring the entire rest mass energy of the dark matter to readily-detectable photons.
At DM Compton wavelengths around or above meter scales,
conversion experiments based on resonant receivers~\cite{Sikivie:1983ip,Krauss:1985ub,Sikivie:1985yu} are a practical
solution, as illustrated by the ADMX experiment~\cite{Asztalos:2009yp,Stern:2016bbw},
and by a range
of ongoing and proposed experiments at similar and lower frequencies
\cite{Lamoreaux:2013koa,Brubaker:2016ktl,Kahn:2016aff,Chaudhuri:2014dla}.
At higher DM masses, filling a large volume with resonant elements,
such as cavities
matching the DM Compton wavelength, becomes difficult.
Consequently, other forms of target
structure that can correct the mismatch between the DM and photon
dispersion relations are more practical.
`Dielectric haloscopes'~\cite{TheMADMAXWorkingGroup:2016hpc,etiger1,etiger2}
provide an example of this idea; a periodic structure of alternating
dielectrics modifies photon propagation in the target volume, enabling
DM-to-photon conversion for DM Compton wavelengths matching the
target periodicity.
The proposed MADMAX experiment~\cite{TheMADMAXWorkingGroup:2016hpc,
MADMAXinterestGroup:2017bgn,Millar:2016cjp} aims
to search for axion DM using this technique, over a mass
range $40 - 400 \, \mu {\rm eV}$.

The main topic of this paper will be extending the dielectric haloscope
concept to higher-than-microwave frequencies. At these shorter
wavelengths, it becomes more difficult to construct and manipulate
individual, wavelength-scale elements. On the other hand, it is possible
to make bulk materials whose optical properties vary on the relevant
scales,  all the way down to ultraviolet wavelengths. 
These `photonic' materials have been used to create many
novel optical devices, such as very high quality cavities and
filters~\cite{photonicbook,Sekoguchi:14,Noda2007,Bushell2017}. The
simplest and most widely used examples are multilayer films, as employed
in optical coatings.

There are a number of reasons why higher-mass bosonic DM is an
attractive target for experimental searches. Practically speaking,
single-photon detection becomes significantly easier at energies
$\gtrsim 0.1 \eV$, corresponding to the energy resolution of
superconducting detectors, as made use of in~\cite{Arvanitaki:2017nhi}.
On the theoretical side, there are ranges of parameter space where
simple early-universe production mechanisms can produce the correct DM
abundance, with couplings below current constraints; in particular,
purely gravitational production during inflation can result in a DM
abundance of light bosons. Such cold bosonic dark matter acts as a
coherent classical-like field, oscillating at a frequency set by its mass $m$ and with
an amplitude set by $m$ and the dark matter density. It is coherent
over times of order $t_{\rm coh}\sim(m v^2)^{-1}$ and lengths of order
$l_{\rm coh}\sim(m v)^{-1}$, where $v\sim10^{-3}$ is the virial velocity
in the galaxy. 

\begin{figure}
	\includegraphics[width=0.9\columnwidth]{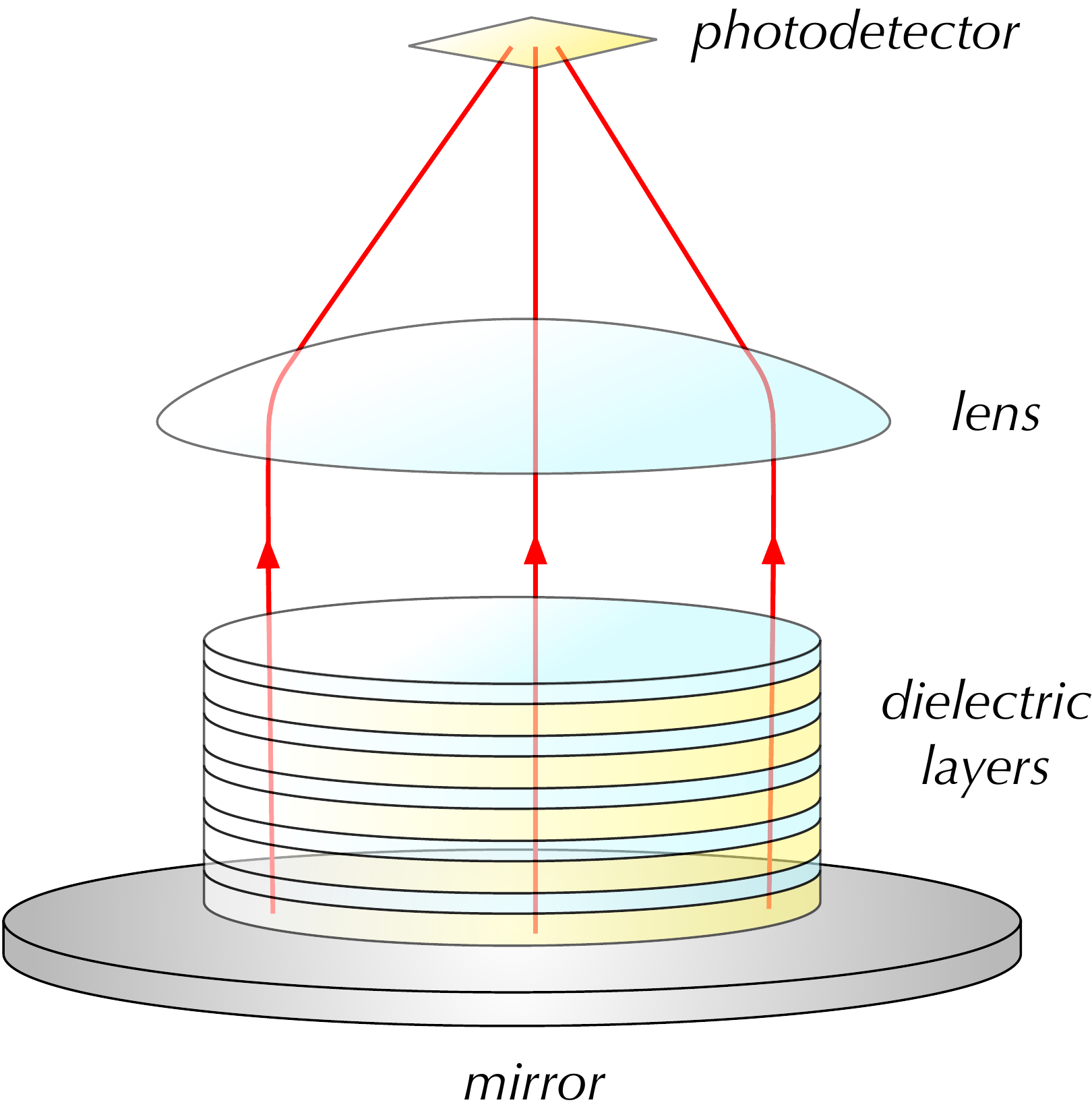}
	\caption{Sketch of our proposed experimental setup. A stack of
	dielectric layers, with alternating indices of refraction, is placed
	on a mirror. In the presence of the right type
	of background DM oscillation (e.g.\ dark photon DM),
	at a frequency corresponding to the inverse
	spacing between the layers, the layers will emit
	photons in their normal direction (shown as red lines). These can be focused onto
	a sensitive, low-noise detector. To detect axion DM with a coupling 
	to photons, a magnetic field should be applied parallel to the layers.
	}
	\label{fig:setup2d}
\end{figure}

In this work we outline an experimental proposal using 
multilayer films, combined with a sensitive photodetector, to search for
bosonic dark matter. Alternating layers of commonly used dielectrics
with different indices of refraction lead to coherent conversion of
dark photon, and, in the presence of an applied magnetic field, axion,
dark matter to photons. The resulting photons emerge in a direction
perpendicular to the layers, and are focused onto a 
detector (Figure~\ref{fig:setup2d}).
These setups have close-to-optimal DM absorption rates (in a DM mass-averaged sense).
Small volumes ($\sim {\rm cm}^3$) of layered material could 
achieve sensitivities several orders of magnitude better than existing
constraints for these dark matter candidates.

Section~\ref{sec:multilayer} discusses the theory of 
DM absorption in layered materials; this can be skipped
by readers more interested in experimental details.
In Section~\ref{sec:setup}, we summarize these theoretical results, discuss concrete, illustrative
examples of how such an experiment might be realized,
and analyze the sensitivity of these setups.
Section~\ref{sec:othercouplings} discusses sensitivity to other forms
of DM. In Section~\ref{sec:dmprod}, we give a brief overview of DM production
mechanisms, and conclude by discussing future extensions and comparing our proposal to other experiments in Section~\ref{sec:discussion}.


\section{Multilayer optical films}
\label{sec:multilayer}

In this Section, we will start by giving a brief overview of
the physics of DM to photon conversion, in the simplest
photonic materials: multilayer films. By the scaling properties
of Maxwell's equations, this is a rescaled version of the
physics of dielectric haloscopes at microwave frequencies,
as derived in depth by various publications from the MADMAX
collaboration~\cite{TheMADMAXWorkingGroup:2016hpc,Millar:2016cjp}. 
Here, we derive the results needed for our experimental configurations
from a slightly different perspective, giving 
some additional physical insight.

\subsection{Axion conversion in layered materials}
\label{sec:halfwave}

DM to photon conversion is an especially attractive experimental
strategy for models in which the DM couples directly to
the EM field --- for example, an axion $a$ with Lagrangian
\begin{align}
	\LL &\supset \frac{1}{2} (\partial_\mu a)^2 - V(a)
-\frac{1}{4} g_{a \gamma \gamma} a F_{\mu \nu} \tilde{F}^{\mu\nu} \nonumber\\
	&=\frac{1}{2} (\partial_\mu a)^2 - V(a) +g_{a \gamma \gamma} a E \cdot B,
\end{align}
where we take the $(+---)$ signature, and use the convention
$\epsilon_{0123} = -1$. Except where indicated, we use natural units
with $c = \hbar = 1$.
By `axion', we will mean a spin-0 particle with (dominantly) odd-parity
couplings to SM states, of which a QCD axion would be a particular example.
An axion generally has a periodic
potential, $V(a) \simeq - m^2 f_a^2 \cos (a / f_a)$, where $f_a$
is the axion's `decay constant', and $m$ is its mass.
The `natural' expectation for the coupling to photons is
that 
$g_{a \gamma \gamma} \simeq \frac{\alpha_{\rm EM}}{2\pi f_a}$,
where $\alpha_{\rm EM}$ is the fine structure constant~\cite{diCortona:2015ldu}.
As discussed in Section~\ref{sec:dmprod}, a dark matter
abundance of these axions could be produced via a number of mechanisms.
We will take such an axion as our prototypical example for the rest
of this section, commenting later on sensitivity to other DM
candidates.

In the presence of the $a F \tilde{F}$ interaction term, the Maxwell equations are
modified to~\cite{Wilczek:1987mv}
\begin{eqnarray}
	\nabla \cdot E &=& \rho - g \nabla a \cdot B,\nonumber\\
	\nabla \times B &=& \partial_t E + J + g (\dot a  B + \nabla a \times E),\nonumber\\
	\nabla \cdot B &=& 0 \quad , \quad \nabla \times E = - \partial_t B,
\end{eqnarray}
where we abbreviate $g_{a\gamma\gamma}$ as
$g$, here and following. In particular, a uniform oscillating
$a$ field has the same effects as an oscillating current density $g \dot
a B$ in the direction of $B$, while other contributions are suppressed
by the DM velocity. Accordingly, the ideal `target' for axion to photon
conversion
is a strong magnetic field; in common with other axion detection
experiments, 
we will use an approximately uniform field from a large magnet.

\begin{figure}
	\includegraphics[width=0.8\columnwidth]{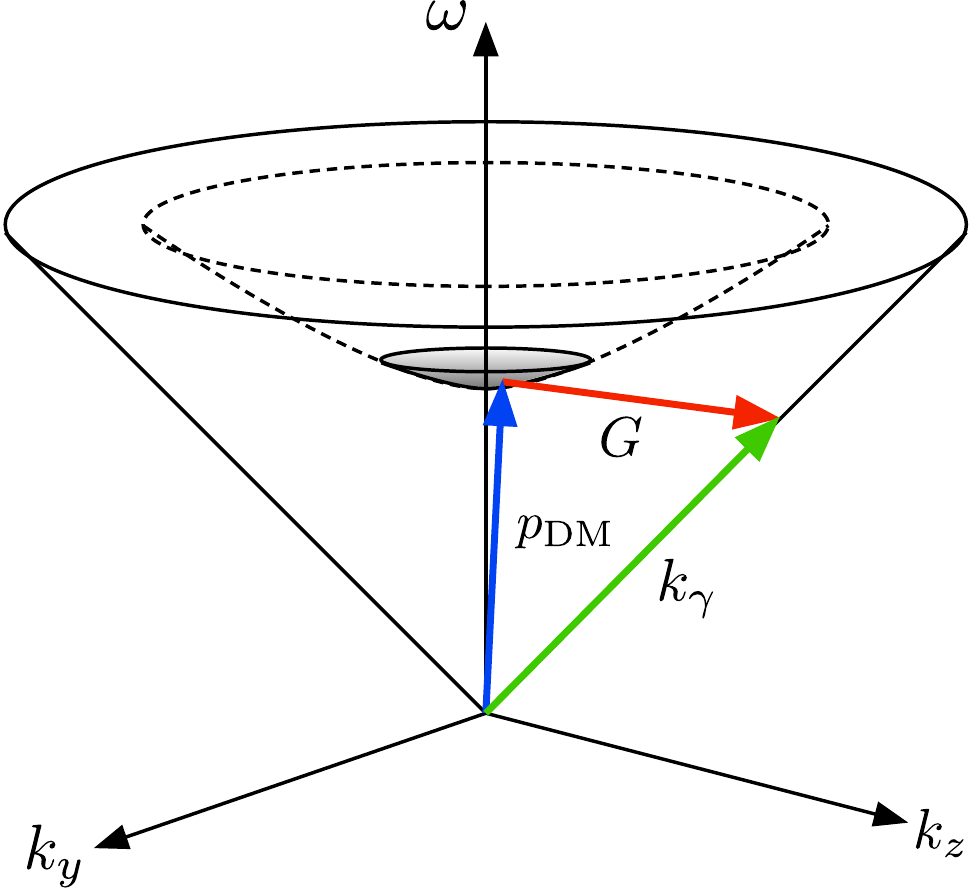}
	\caption{Schematic illustration of DM with 4-momentum $p_{\rm DM}$ being converted to
	a photon with momentum $k_\gamma$. The dotted line indicates the mass shell
	of the DM particle; since the DM is non-relativistic, a given quantum
	will have small velocity, corresponding to the shaded region.
	For this illustration, we have assumed that photons have a linear dispersion
	in the target material (see Figure~\ref{fig:bandgap} for how
	this can be modified).
	In periodic materials, conservation of pseudo-momentum inside
	the material requires that $k_\gamma - p_{\rm DM} \equiv G$ is a reciprocal
	lattice vector of the medium.}
	\label{fig:pm1}
\end{figure}

The remaining difficulty in accomplishing $a-\gamma$ conversion
is achieving a setup in which the interaction does not cancel
out when integrated over the target. For resonant cavity experiments
such as ADMX, modes above the few lowest-lying ones will have small
overlap with the effective current $g \dot a B$.
In target materials with small-scale periodicity, another formulation
is that DM-photon conversion process must conserve (pseudo)momentum,
up to the material's reciprocal lattice vectors, analogously to Bragg
scattering. As Figure~\ref{fig:pm1} schematically illustrates, this means that
the target must have structure on scales set by the Compton
wavelength of the DM.

A simple way to realize the appropriate structure is with photonic
materials, which have spatially non-uniform optical properties.\footnote{Another possibility would be to use a spatially non-uniform magnetic field,
as per the ORPHEUS experiment~\cite{Rybka:2014cya}, but this is harder
to implement with a multi-tesla field.} Figure~\ref{fig:setup2d} shows the schematic structure of such a detector
using a 1D photonic material: a set of dielectric layers with alternating
permittivities. As illustrated by Figure~\ref{fig:pm1},
most of the momentum of the emitted photons `comes from' the periodicity
of the material, so they are emitted in a tight cone around a particular angle.
They can then be focused down onto a small, sensitive detector.
This is precisely the experimental setup proposed, at microwave
frequencies, by the MADMAX collaboration~\cite{TheMADMAXWorkingGroup:2016hpc}.

\begin{figure}
	\includegraphics[width=\columnwidth]{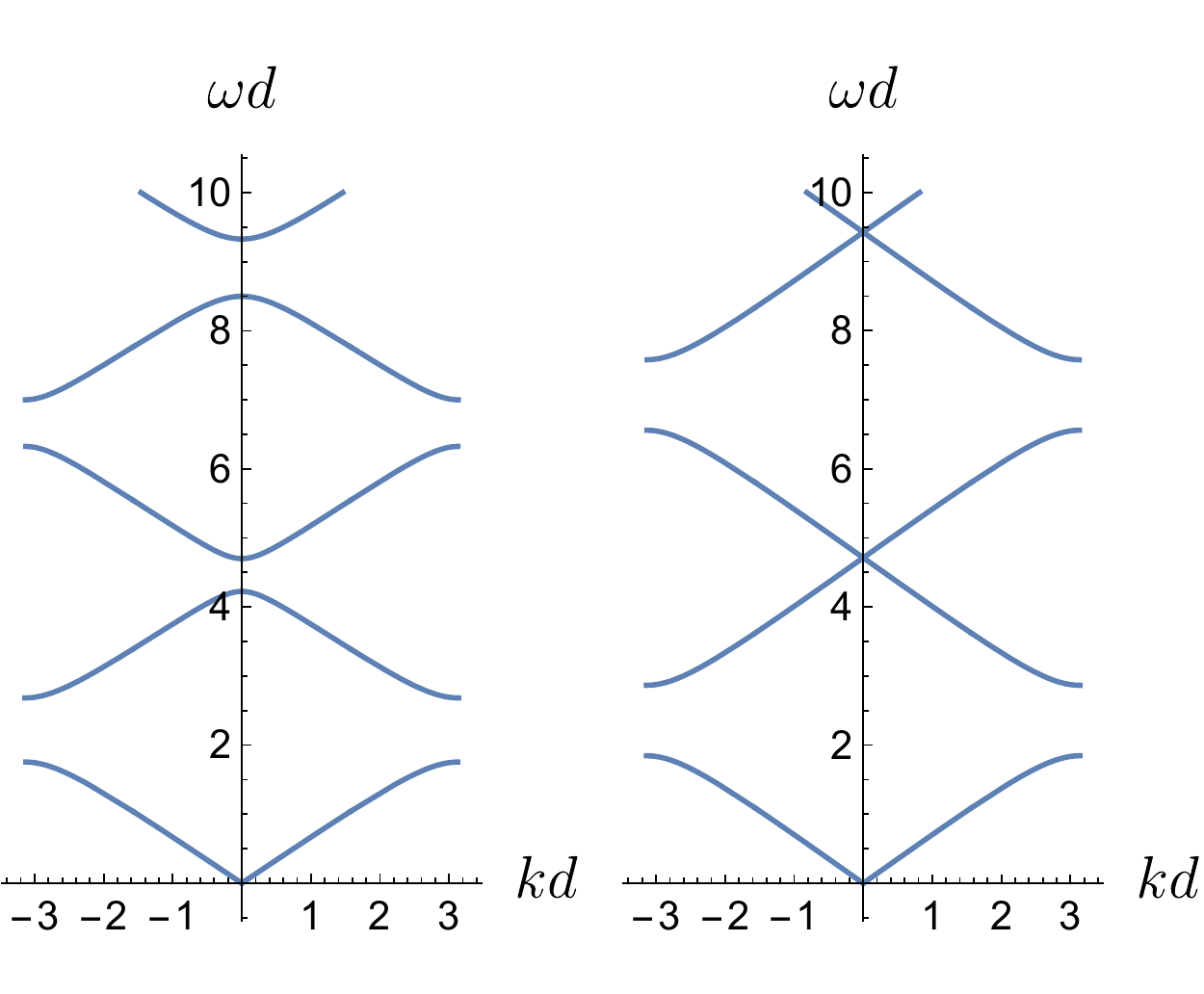}
	\caption{\emph{Left-hand plot:} dispersion relation for photon
	propagation in a periodic medium with alternating layers of refractive
	indices $n_1 = 1, n_2 = 2$, and widths $d_1 = 1, d_2 = 0.7$ 
	(in arbitrary units), where $d = d_1 + d_2$ is the periodicity.
	The momentum $k$ is taken to be in the same direction as the material
	periodicity.
	\emph{Right-hand plot:} as for the left-hand plot,
	but with $d_1 = 1, d_2 = 0.5$. As discussed in Section~\ref{sec:halfwave},
	this configuration has no band-gaps around the $k = 0$ points.}
	\label{fig:bandgap}
\end{figure}

The simplest situations to analyze correspond to periodic layered structures.
Photon modes in an infinitely extended
periodic medium are Bloch modes; if the layers are uniform
in the $x,y$ plane, then
\begin{equation}
	E(r) = e^{i k \cdot r} u_{k} (z),
\end{equation}
with $u_k(z + d) = u_k(z)$, where $d$ is the periodicity in $z$ of
the material. Taking $k_z$ to lie within the first Brillouin zone,
$(-\pi/d,\pi/d)$, we can label different modes at the same $k_z$
by band number. Figure~\ref{fig:bandgap} illustrates part of this 
band structure for some simple periodic materials.
The modes in a large
but finite stack will be similar to those in the infinite case.
As noted above, the DM momentum is small compared to its
mass, so we are interested the modes for which $k_\perp$ is small
compared to $d^{-1}$.
In particular, a DM mode with frequency $\omega$, and small momentum $k$,
can convert to a photon of frequency $\omega$ if the Bloch mode at
that frequency has Bloch momentum close enough to $k$.

The DM absorption rate of any 1D stack configuration can
be calculated via transfer matrices, as presented in~\cite{Millar:2016cjp}.
However, in terms of developing a physical picture, it can
be useful to present the calculation, at least around the $k=0$ points,
in terms of the unforced photon modes in the layers
(this corresponds to the `overlap integral' calculations
of~\cite{TheMADMAXWorkingGroup:2016hpc,Millar:2016cjp}).
The simplest-to-analyse periodic configurations have
layers of alternating refractive indices
$n_1, n_2$, with thicknesses $d_1, d_2$ such that their
phase depths are equal, $n_1 d_1 = n_2 d_2 = \pi/\omega$.

We start by ignoring the DM velocity,
and treating the DM field as a classical background
oscillating at frequency $\omega$,
with $a(t) = a_0 \sin \omega t$ (treating the DM field as classical
will always be a good approximation
in the regimes we consider).
This induces an electric field $E(x,t)$
in the material, resulting in an instantaneous absorbed power of
\begin{equation}
P_{\rm int} \simeq \int dV E \cdot \partial_t (g a B_0),
\end{equation}
where $B_0$ is the background magnetic field
(this is the instantaneous energy flow from the DM field to the SM target).

\begin{figure}
	\includegraphics[width=0.8\columnwidth]{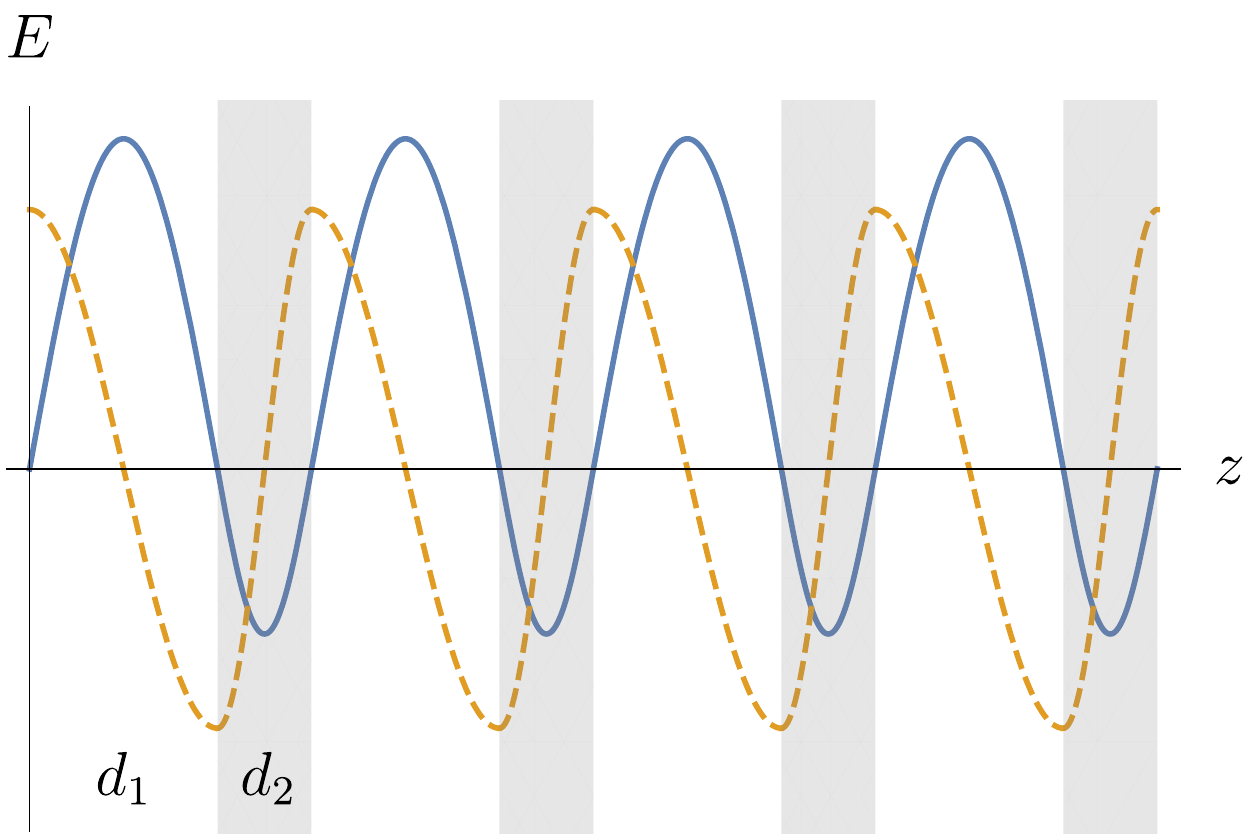}
	\caption{Electric field for the second-band modes with
	Bloch momentum $k=0$, in infinite periodic alternating layers with refractive
	indices $n_1 = 1, n_2 = 2$, 
	and thicknesses $d_1 = 1, d_2 = 0.5$ (as per the right-hand plot
	of Figure~\ref{fig:bandgap}). The $y$ axis denotes
	the amplitude of the electric field, which is transverse
	to the layers. 
	The solid-line mode, which has  non-zero
	$\int dz \, E$ over a period, is the one excited by a background
	DM field. The out-of-phase mode (dashed line) is not excited.}
	\label{fig:emodes}
\end{figure}

At points in the cycle when $a = 0$, the electric field is continuous
at the layer interfaces. The EM fields, at these
moments, must correspond to those of a free photon mode.
The out-of-phase component of the electric field does not 
contribute to the absorbed power, so we can calculate
this taking only the free mode into account.
For a half-wave stack with refractive indices $n_1, n_2$, and layer
thicknesses $d_i = \pi / (n_i \omega)$, there is a mode with electric
field (parallel to the layers)
\begin{equation}
	E = \begin{cases}
		\frac{1}{n_1} \sin (k_1 z) E_0 \cos\omega t & (0 \le z \le d_1)
		\\ \frac{1}{n_2} \sin (k_2 (z - d_1)) E_0 \cos\omega t & (d_1 \le z \le d_1 + d_2),
	\end{cases}
\end{equation}
as illustrated in Figure~\ref{fig:emodes} (there is also
the out-of-phase mode, for which $E$ averages to zero across a period).
If $E_0$ is the amplitude in this mode excited by the DM oscillation,
then the instantaneous absorbed power is
\begin{equation}
	P_{\rm int} \simeq N A \left( \frac{1}{n_1^2} - \frac{1}{n_2^2}\right)	E_0 g a_0 B_0 \cos^2 \omega t,
\end{equation}
where $A$ is the cross-sectional area of the layers,
assuming a uniform background magnetic field $B_0$ parallel to
$E_0$.\footnote{There is an ambiguity
coming from the integral outside
the stack, which cancels over a period, but this is only a $\sim 1/N$
fractional contribution~\cite{Millar:2016cjp}.} Writing the power lost from the stack as $P_{\rm loss}
= \omega U_{\rm osc} / Q$, where 
\begin{equation}
	U_{\rm osc} = \int dV \epsilon \langle |E|^2 \rangle  = \frac{N A}{\omega} \frac{1}{2}
	\left (\frac{1}{n_1} + \frac{1}{n_2}\right) E_0^2
\end{equation}
is the energy stored in this mode~\cite{photonicbook}, we 
obtain a cycle-averaged power loss of
\begin{align}
	\langle P_{\rm abs} \rangle 
	&= \frac{\langle P_{\rm int} \rangle ^2}{\langle P_{\rm loss} \rangle}
	\simeq \frac{Q}{\omega}
	\frac{\langle P_{\rm int} \rangle ^2}{\langle U_{\rm osc} \rangle}
	\nonumber\\
	&=\frac{1}{2}  (g a_0 B_0)^2 Q A N 
	\left(\frac{1}{n_1} + \frac{1}{n_2}\right) \left(\frac{1}{n_2} - \frac{1}{n_1}\right)^2
	\nonumber\\
	&= g^2 B_0^2 \frac{\rho_{\rm DM}}{m^2} Q A N 
	\left(\frac{1}{n_1} + \frac{1}{n_2}\right) \left(\frac{1}{n_2} - \frac{1}{n_1}\right)^2,
	\label{eq:pabsq}
\end{align}
where $\rho_{DM} \simeq 0.3 \GeV/\cm^3$ is the local dark matter density and $m$ is the axion mass.
This corresponds to solving for $E_0$ such that $\langle P_{\rm int} \rangle
= \langle P_{\rm loss} \rangle$.
The `quality factor' $Q$ depends on the stack's surroundings.
For an `open cavity' setup corresponding to a stack of layers in air 
(i.e.\ none of the photons emitted by the layers are reflected back),
the power loss is proportional to the area of the end-caps, so
$Q \propto N$. However, the precise value depends on the form of the
stack-air interfaces. 
For example, if the left-hand end of the stack were an $n_1$ layer,
and the right-hand end an $n_2$ layer, as in Figure~\ref{fig:emodes}, then
$Q = 2 N \left(1/n_1 + 1/n_2\right)$. 

\subsubsection{Other configurations}\label{sec:otherconf}

The `half-wave stack' configurations considered above
have many useful special properties.
For generic layer profiles, there are `bandgaps' around
each $k = 0$ point, as illustrated in Figure~\ref{fig:bandgap}.
At frequencies within the bandgap,
there are no periodic mode solutions --- for a finite set of layers,
incident modes at these frequencies are exponentially attenuated
inside the material. Since the converted photons
produced by the DM must be close to these bandgap edges, this can lead
to complicated behavior (in particular, to very narrow mass range coverage).
However, the bandgaps around the $k=0$ points for half-wave stacks
vanish (right-hand panel of Figure~\ref{fig:bandgap}).
A related advantage of this configuration is that such a stack has
high transparency at frequencies close to zero Bloch momentum.
Consequently, it is easy for photons produced by one stack
to pass through another at a nearby frequency (Section~\ref{sec:chirp}).\footnote{These
configurations, first considered in~\cite{Jaeckel:2013eha}, are referred to by the MADMAX collaboration as `transparent mode'~\cite{Millar:2016cjp}.
}

Another point is that, 
for a stack terminated on one end by a mirror (as per Figure~\ref{fig:setup2d}),
the mirror's boundary condition means that there is only one free mode
at a given frequency, which is a standing wave. The absorbed power will
then be controlled by the $E$ integral of this mode. However, while this
can be very small at the central frequency, if the mirror selects
the `wrong' Bloch mode, we can go to a close-by frequency, $\delta \omega \sim \omega/N$, and recover an order-$N$ overlap. If the mirror
selects out the correct Bloch mode, e.g.\ if the left-hand side of Figure~\ref{fig:emodes}
were replaced by a mirror, then equation~\ref{eq:pabsq} applies.
For a stack-air interface at an $n_2$ layer, $Q =  4 N (1/n_1 + 1/n_2)$.

It is simple to modify $Q$ further by placing (partially) reflecting
surfaces (e.g.\ other dielectric layers) at the ends of a stack;
effectively, placing it inside a cavity.
However, as Section~\ref{sec:freqav} shows, increased $Q$ is always compensated
for by decreased mass range coverage (though this can still be helpful in
terms of signal discrimination and signal-to-noise).

\subsubsection{DM velocity distribution}
\label{sec:vel}

The DM in our galaxy is expected to have a virialized velocity
distribution, with typical velocity $v \sim 10^{-3}$.
Consequently, the DM field has a coherence length 
$l_{\rm coh} \sim 1/(v m)$.
This means that photon emission from points further apart than $l_{\rm coh}$ adds
incoherently, when averaged over long times. Thus, stacks
of $N \gtrsim v^{-1} \sim 10^3$ periods will no longer
have peak conversion power $\sim N^2$.
This can be verified by explicit computations,
as in~\cite{Millar:2017eoc}~\footnote{Note that the
expressions and plots shown in~\cite{Millar:2017eoc} are for emission
from only one end of a (non-mirror-backed) stack; the deviation
of these quantities from their zero-velocity values can be significantly
greater than the deviation in the overall converted power.}.
However, as we discuss in Section~\ref{sec:freqav},
the converted power averaged over DM masses is almost unaffected by
the velocity distribution.

Another effect of the DM velocity distribution 
is on the angular distribution of converted photons.
Since the transverse momentum of a converted
photon is the same as that of the DM quantum, the emitted
photons will be distributed within a cone of opening angle $\sim 10^{-3}$
around the $z$ axis. This is important in terms of focusing the emitted
photons (Section~\ref{sec:photondet}), and could potentially be used
post-discovery to determine the velocity distribution of the DM~\cite{Jaeckel:2013sqa,Jaeckel:2015kea,Arvanitaki:2017nhi}.
The DM velocity distribution also affects the frequency spectrum
of converted photons, and (in a non-mirror-backed stack) the backwards
vs forwards emission rates~\cite{Knirck:2018knd}; if a signal were seen,
these could be checked against each other for consistency with a
DM signal origin.

\subsection{Dark photon conversion}

In addition to axions, light vector DM is another natural
candidate for dielectric haloscope searches. A `dark photon' coupled to
the SM through kinetic mixing with the photon has an unusually large
window of open parameter space, since plasma effects suppress its
emission from stars~\cite{An:2013yfc,Hardy:2016kme,Chang:2016ntp}, and
it does not mediate long-range forces between neutral matter. Also, unlike
spin-0 DM candidates, vector DM has a polarization direction, and can
convert to transverse photons even in a homogeneous target without
velocity suppression.
As discussed in Section~\ref{sec:dmprod}, a dark matter abundance
of such vectors can naturally be produced in the early universe via inflation.

Suppose that dark matter consists of a dark photon $A'$, with
\begin{equation}
	\LL \supset -\frac{1}{4} F_{\mu\nu} F^{\mu\nu}
	- \frac{1}{4} F'_{\mu\nu} F'^{\mu\nu} + \frac{1}{2} m^2 A'^2
	+ J^{\mu}_{\rm EM}(A_{\mu} + \kappa A'_{\mu});
\end{equation}
this is equivalent, after field redefinition,
to the usual `kinetic mixing' interaction $-\frac{1}{2}\kappa F_{\mu\nu}
F'^{\mu\nu}$. Solving for the fields in a uniform dielectric, an oscillatory $A_0'$ background field
 induces a corresponding Standard Model (visible) electric field $E_{\rm vis} = - \kappa \frac{\chi}{\epsilon} E_0'$,
where $E_0' = \partial_t A_0'$ is the dark electric field, and $\chi = \epsilon-1$ is the polarizability~\cite{Jaeckel:2013eha}. This is in comparison
to the axion-induced electric field $E_{\rm vis,a} = - g a B_0 / \epsilon$~\cite{Millar:2016cjp}.
Since the boundary conditions at a dielectric interface are the same in both cases,
the dark-photon induced power (per unit area, from a single interface) is given by
\begin{equation}
	(E_{\rm vis}^1 - E_{\rm vis}^2)_{\parallel}
	= \left(\frac{\chi_1}{\epsilon_1}
	- \frac{\chi_2}{\epsilon_2}\right) \kappa (E_0')_{\parallel}
	= \left(\frac{1}{\epsilon_2}
	- \frac{1}{\epsilon_1}\right) \kappa (E_0')_{\parallel},
\end{equation}
where $\parallel$ denotes the part parallel to the interface.
Thus, the conversion rate for a dark photon amplitude $E_0'$
is equivalent to that for an axion amplitude $a_0$ with $\kappa (E_0')_{\parallel} = 
g a_0 B_0$. For a half-wave stack, 
\begin{equation}
	\langle P_{\rm abs} \rangle  
	= \kappa^2 \sin^2\theta \rho_{\rm DM} Q A N 
	\left(\frac{1}{n_1} + \frac{1}{n_2}\right) \left(\frac{1}{n_2} - \frac{1}{n_1}\right)^2,
	\label{eq:dppower}
\end{equation}
where $\theta$ is the angle of $E_0'$ from the layer normals.
This angle will be approximately constant over the coherence timescale of the
DM field, $t_{\rm coh} \sim m^{-1} v^{-2}$, and will vary over longer timescales;
for an isotropic DM velocity distribution, the long-time average
of $\sin^2 \theta$ is $2/3$. 
One difference between the dark photon and axion cases is that, for the latter,
the polarization of the emitted photons is set by the $B_0$ field,
while for a dark photon, it is set by the DM itself.

From equation~\ref{eq:dppower}, the dark photon to photon
conversion rate is independent of $m$, for a given target volume 
(since the length of an $N$-period stack is $\propto m^{-1}$).
This may naively be worrying, 
since the dark photon
has to decouple from the SM in the $m \rightarrow 0$ limit. For 
cavity experiments such as those proposed in~\cite{Chaudhuri:2014dla}, 
this manifests itself as a $(m L)^2$ suppression of the conversion
rate, where $L$ is the scale of the experiment, for $m L \ll 1$. However,
in our case, the width of the layers is set by $m^{-1}$, so when our expressions
are valid, the experiment is automatically much larger than
the DM Compton wavelength.


\subsection{Frequency-averaged power absorption}
\label{sec:freqav}

\begin{figure}
	\includegraphics[width=0.9\columnwidth]{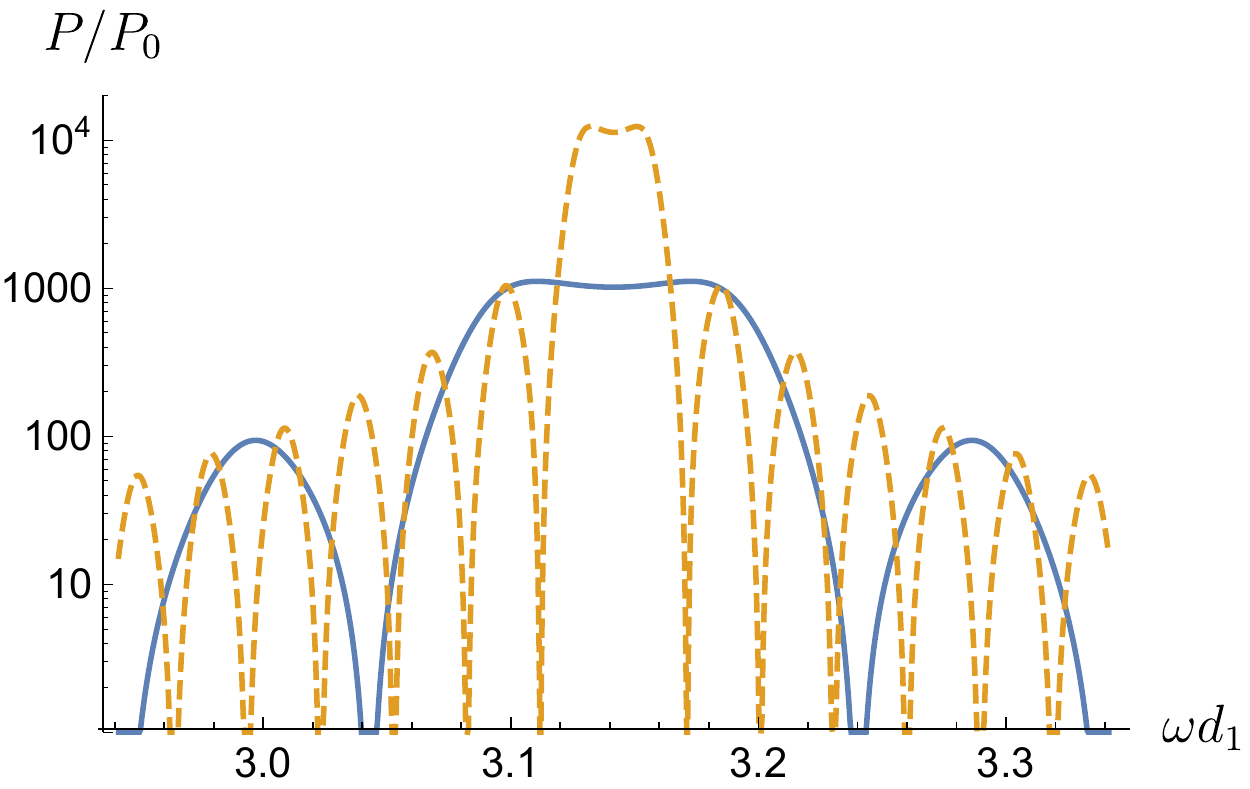}
	\caption{Power absorbed from spatially-uniform DM
	oscillation, as a function of dark matter frequency $\omega$,
	for a half-wave stack of layers with refractive indices
	$n_1 = 1, n_2 = 2$, showing results for 30 (solid blue) and 100 (dashed orange)
	periods. As discussed in the text,
	the peak conversion power increases as $N^2$, where $N$ is
	the number of layers, but the mass range over which this
	holds decreases as $1/N$. 
	The reference power is defined by $P_0 \equiv g^2 B_0^2 \rho_{\rm DM} \omega^{-2}$, for axion DM in a uniform background
	magnetic field $B_0$.}
	\label{fig:pcomp}
\end{figure}

Considering a spatially-uniform DM field oscillation to begin with,
a natural expectation is that the fractional range in frequencies
over which we get $\sim N^2$ converted power (eqns.~\eqref{eq:pabsq},\eqref{eq:dppower}) is $\sim 1/Q$.
For example, a change $\delta \omega / \omega \sim 1/N$ leads to a
$\sim 1$ change in phase across $N$ layers, destroying the coherent addition.
Figure~\ref{fig:pcomp} shows this scaling for example values of $N$.
Thus, while the peak conversion power is $\sim N^2$, the averaged power
over an $\OO(1)$ range in frequencies is $\sim N$, and is not coherently
enhanced.
If we do not know the DM mass, which could a priori be anywhere
in a wide interval, then the frequency-averaged power controls
how fast we can scan over a range of masses (for a low-background experiment).

We can see the $\sim N$ scaling in more generality from an `impulse response' argument
(for clarity, we take the example of axion DM here; the same arguments
apply to a dark photon).
For a target consisting of a set of thin interfaces between uniform layers,
a sufficiently short `pulse' $a(t)$ of the DM field must, immediately after
its arrival, result in only local field disturbances around the interfaces:
 influences have not had time to propagate further, and a uniform
lossless dielectric does not absorb any energy. In the linear regime, the response of the target
to a superposition of DM signals is the superposition of the responses
to each signal; accordingly, we can decompose the pulse into sinusoidal components.
Analyzing the response of a single interface $i$ to a sinusoidal DM oscillation,
we find that the cycle-averaged power absorbed, per unit area,
is
\begin{equation}
	\frac{\langle P_{\rm abs}\rangle_i}{A}  = \frac{(g a_0 B_0)^2}{2}
	\left(\frac{1}{n_1} + \frac{1}{n_2}\right) \left(\frac{1}{n_2} - \frac{1}{n_1}\right)^2,
	\label{eq:player}
\end{equation}
where $n_1$ and $n_2$ are the refractive indices on either side
of the interface~\cite{Millar:2016cjp}.
So, for a pulse $a(t)$ shorter than $\sim \min \{1/(n_i d_i)\}$, for which the power
absorbed should sum incoherently across different interfaces, 
the total energy absorbed (per unit area) is
\begin{eqnarray}
	\frac{U_{\rm abs}}{A} &\simeq& \left( \int d\omega \, |\tilde a(\omega) |^2  \right) \times \nonumber\\
	&&\frac{g^2 B_0^2}{2}
	\sum_{i}
	\left(\frac{1}{n_1} + \frac{1}{n_2}\right) \left(\frac{1}{n_2} - \frac{1}{n_1}\right)^2,
	\label{eq:uabs}
\end{eqnarray}
where $\tilde a(\omega)$ is the Fourier transform of the pulse $a(t)$,
and the sum runs over interfaces $i$.
If there is refractive index structure on significantly smaller scales,
that can be treated as an effective medium of averaged
refractive index.
Since, in the linear regime, the power absorbed depends only
on the power spectrum of the DM signal, and not on the relative
phases of different frequency components, we can infer that
for a general DM signal, with power spectral density $S_{aa}(\omega)$, the
long-time average power absorbed, per unit area, is
\begin{align}
	\frac{\langle P_{\rm abs}\rangle}{A} \simeq &\left(\int d\omega \, S_{aa}(\omega) \right) \times \nonumber\\
	&\frac{g^2 B_0^2}{2}
	\sum_{i}
	\left(\frac{1}{n_1} + \frac{1}{n_2}\right) \left(\frac{1}{n_2} - \frac{1}{n_1}\right)^2,
\end{align}
if $S_{aa}(\omega)$ is a broad distribution over the appropriate frequency
range, and the refractive indices do not appreciably change
over this frequency range.\footnote{This corresponds to
the `area law' derived in~\cite{Millar:2016cjp}, but extends it by giving
an explicit expression for the frequency-averaged power in terms of the
refractive indices of the layers.} For periodic layer spacings, we can
be more specific;
in the case of a half-wave stack, the converted power is a periodic
function of frequency (for given DM amplitude $a_0$) with period $2 \omega_0$,
where $\omega_0$ is the half-wave frequency. Consequently, if we 
consider a flat power spectral density $S_{aa}(\omega)$, we find that
the frequency-averaged conversion power is
\begin{equation}
	P_{\rm av} = (g a_0 B_0)^2 A N  
		\left(\frac{1}{n_1} + \frac{1}{n_2}\right)
		\left(\frac{1}{n_2} - \frac{1}{n_1}\right)^2,
			\label{eq:pavax}
\end{equation}
where we  take the average to be over $0 < \omega < 2 \omega_0$.
For a half-wave stack, most of this power is in the peak,
at frequencies within $\sim 1/Q$ of $\omega_0$, as illustrated in
Figure~\ref{fig:pcomp}.

\begin{figure}
	\includegraphics[width=0.9\columnwidth]{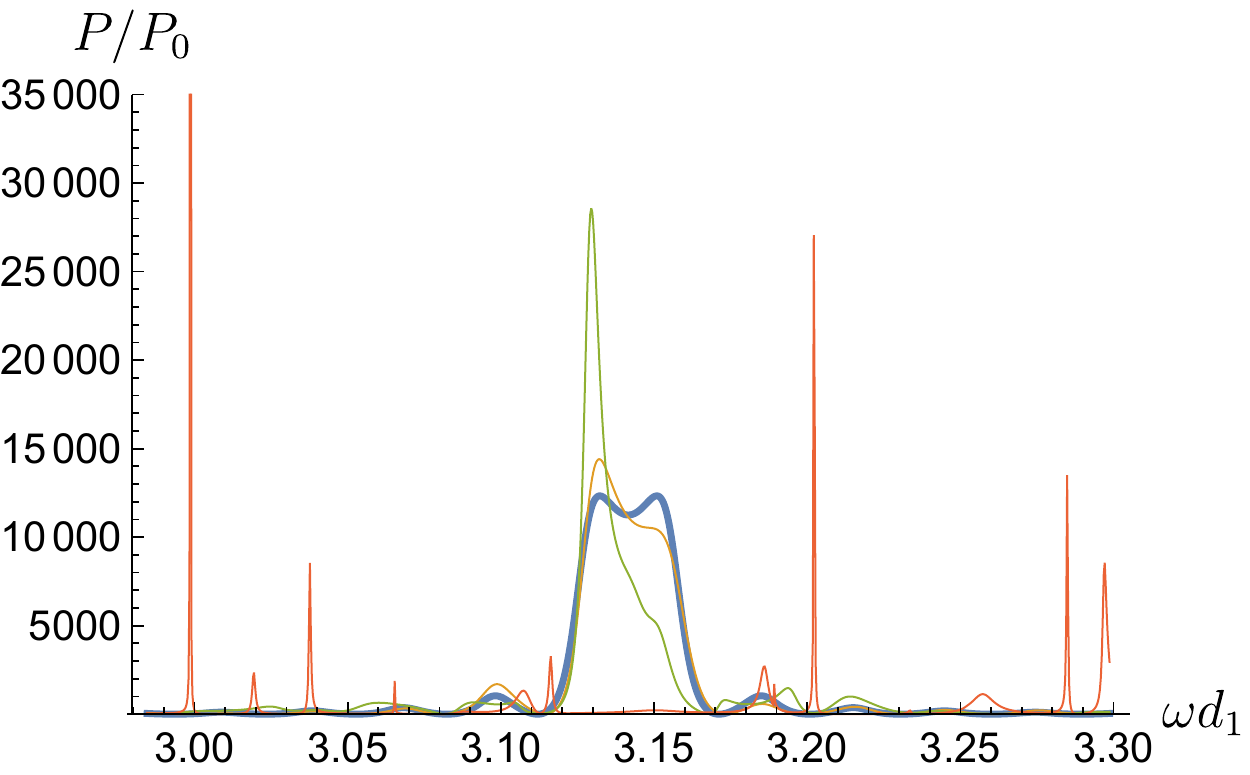}
	\caption{Blue (thick) curve: power absorbed from spatially uniform DM oscillation at a single
	frequency $\omega$, for a 100-period half-wave stack with refractive indices
	$n_1 = 1, n_2 = 2$.
	The (thin) orange, green and red curves show the effect of introducing
	(uncorrelated) random thickness differences in the layers, with
	fractional deviation
	0.01, 0.05, and 0.2. This illustrates that, when the fractional deviations
	are small compared to $\sim 1/\sqrt{2 N}$, the effect on the frequency profile is
	small, whereas for larger deviations, the frequency profile changes
	completely, becoming spiky.
	The reference power is defined by $P_0 \equiv g^2 B_0^2 \rho_{\rm DM} \omega^{-2}$, for axion DM in a uniform background
	magnetic field $B_0$.}
	\label{fig:rand1}
\end{figure}

If we do not know the DM mass, then since the DM is non-relativistic,
this corresponds to not knowing the oscillation frequency of $a$.
Suppose that we have some number of different experimental configurations
which, in combination, provide sensitivity over a $\OO(1)$ range
of $m$. Then, the total energy converted by all of
the experiments, at the $m$ for which this is lowest, is at most
the average energy converted over the whole range of $m$.
This corresponds to
taking $S_{aa}(\omega)$ approximately constant over the given frequency range,
and summing across all of the experimental configurations.
Thus, the minimum time-averaged powered converted by the set of 
configurations is at most
\begin{equation}
	\langle P_{\rm min} \rangle \lesssim \frac{\rho_{\rm DM}}{m_{\rm max}^2}
	g^2 B_0^2
	\sum_{s} A_s
	\sum_{i}
	\left(\frac{1}{n_1} + \frac{1}{n_2}\right) \left(\frac{1}{n_2} - \frac{1}{n_1}\right)^2,
\end{equation}
where $m_{\rm max}$ is the upper end of the mass range covered,
and $s$ runs over the different stacks.
For a dark photon, the corresponding expression has $g^2 B_0^2 / m_{\rm max}^2$
replaced by $\frac{2}{3} \kappa^2$.
If the power converted as a function of mass is too `spiky',
this bound maybe be difficult to attain; for example, Figure~\ref{fig:rand1} shows
how random layer spacings result in sharp peaks at a fairly
random set of frequencies, despite having similar frequency-averaged power
(see Section~\ref{sec:tol}). However, simple frequency profiles,
such as that obtained from a half-wave stack, can easily be added
together to obtain smooth coverage over an order-1 mass range,
as illustrated in Figure~\ref{fig:layercomb}.

If an experiment is not background free,
the frequency-averaged conversion power will not be the only relevant
quantity. Spending shorter times searching more, but narrower,
frequency ranges will result in the same total number
of signal photons produced over the whole lifetime of the experiment, but
these all come within a shorter time,
improving the signal to noise.
However, as we will review in Section~\ref{sec:setup}, 
close to background-free photon detectors are possible over
most of the frequency range we are considering.
Consequently, in searching for a DM signal of unknown mass over a broad
range, frequency-averaged power is a useful figure of merit.
The simplest way to cover a large mass range is either to construct a set of stacks,
each covering a small part of the range, or to construct
a single stack with segments of different periodicities
(a `chirped stack').
In the latter case, the transparency properties of the half-wave
stack configuration discussed above are useful, as discussed
in Section~\ref{sec:chirp}. \footnote{MADMAX ~\cite{TheMADMAXWorkingGroup:2016hpc,Millar:2016cjp}
addresses the problem of achieving narrow mass coverage,
$\delta \omega/\omega \ll 1/N$, using a small number ($N \sim$ tens)
of slabs. This bandwidth is then scanned by physically repositioning
the slabs.
Their experiment is limited by thermal backgrounds, so narrow mass
coverage improves the signal to noise. In our case, where
such tuning may be difficult, but many more layers fit into a
reasonable volume, the problem instead becomes covering
a broad mass range without having to construct an enormous
number of stacks; hence our emphasis on the half-wave stack.
}

The above results apply for spatially-uniform DM oscillations,
i.e.\ $v_{\rm DM} = 0$.
As mentioned in Section~\ref{sec:vel}, the DM velocity distribution,
which is spread over a range $\delta v \sim 10^{-3}$, has a significant
effect on the converted power when the layers are spaced over $\gtrsim
10^3$ wavelengths. 
However, by the same impulse response arguments
as above, the frequency-averaged converted power is a local property
for each interface. Since, for $\delta v \sim 10^{-3} \ll 1$,
the scale over which the DM field varies is much larger
than a signal photon wavelength, then if we average
over a broad range of DM masses, the total power converted by all
of the interfaces is still given by equation~\ref{eq:pavax}, to a good
approximation.
If we look at the total converted power as a function of DM mass,
then the effect of the velocity distribution is
to spread this out (by $\sim \delta v$ in fractional mass range),
while almost preserving its mass-averaged value.

\subsection{Tolerances}
\label{sec:tol}

As discussed above, periodic layers have the advantage, compared to more random
configurations, that their converted power is a smoother function
of frequency (i.e. of DM mass), especially in the case of a half-wave stack.
However, imperfections in the manufacturing process will result in some unintended variation in layer properties.

Considering flat, parallel interfaces, if the total accumulated phase error across all of the layers in a half-wave
stack is $\gtrsim 1$, then we expect the 
frequency profile to change significantly (though, as per above, the frequency-integrated
power will stay approximately constant). Such deviations could arise from a
combination of modified layer thicknesses and refractive indices.
If deviations in different layers are uncorrelated, per-layer
fractional deviations of up to $\sim 1/\sqrt{2 N}$ will not significantly
affect the profile (as is also derived in~\cite{Millar:2016cjp}). This is illustrated in Figure~\ref{fig:rand1}, which
also shows an example of the highly modified `spiky' profile resulting
from larger random deviations.

If we allow position-dependent thickness and/or index variations,
resulting in non-planar layers, then the same $\lesssim 1/\sqrt{2 N}$
condition on uncorrelated fractional deviations is required at each position.
This also ensures that the emitted photons are kept within the cone
set by the DM velocity, with opening angle $\sim 10^{-3}$, as required
for optimum focussing; while the layers may be `bumpy', the bumps
have sub-wavelength height, and diffraction ensures that the overall emission
is still collimated. As an additional point, we do not necessarily require that
the fractional deviation condition applies strictly over the whole
area of the stack. Emission from areas of the stack separated by more than a DM
coherence length adds incoherently --- therefore, to avoid
a spiky frequency profile or uncollimated emission, we only
need the fractional deviation condition to hold within these small areas.
If layer thicknesses and/or indices change smoothly over larger
distance scales, then different cross-sectional pieces of the stack
effectively have different central frequencies, and add incoherently
(as illustrated in the blue curve of Figure~\ref{fig:layercomb}).

The complicated frequency profiles of randomly-spaced layers
occur because of their effect on photon propagation.
In a random medium, instead of the definite bandgaps of a periodic medium, the
 frequency range corresponding to the inverse scale
of variation becomes a `pseudogap', in which photons propagate
diffusively~\cite{Anderson1985,John:1987zz}. The very long `effective path length'
for a photon to escape the layers means that a very small change in frequency
can significantly change the photon mode, leading to a very quickly-varying
absorption rate with frequency.

So far, we have used the approximation of a lossless dielectric.
However, a real material will absorb some of the light passing through
it. Considering a photon mode in the layers, absorption
will become important when the damping rate is $\gtrsim \omega / Q$,
where $Q$ is the mode's quality factor. The half-wave stacks
considered above have $Q \sim N$, so if the imaginary part of
the refractive index is $\lesssim 1/N$, then absorption will not be important.
Configurations with narrower frequency peaks will be correspondingly more affected.

\subsection{Chirped stacks}
\label{sec:chirp}

\begin{figure}
	\includegraphics[width=0.9\columnwidth]{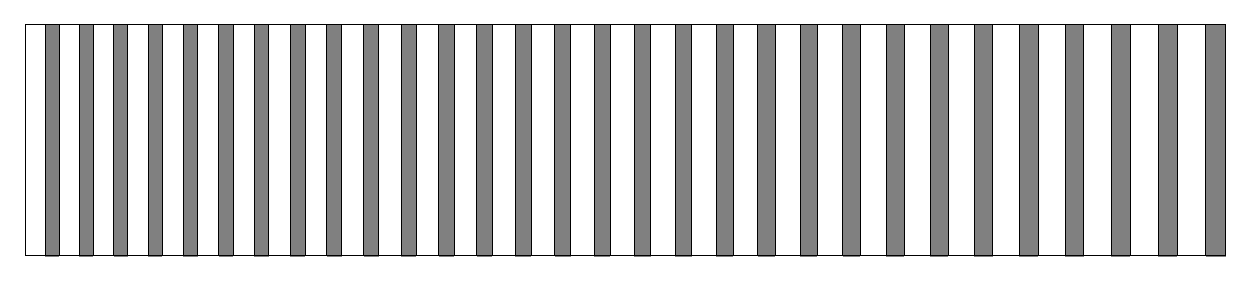}
	\caption{Illustration of a `chirped stack' configuration, in which
		the layer spacings are gradually changed across the length of
		the stack. This example has 30 periods, with refractive indices
		$n_1 = 1, n_2 = 1.46$ (analogous to alternating gas/silica
		layers --- see Section~\ref{sec:dielectrics}).
		The phase depths of adjacent layers are very close to equal,
		so that it locally approximates a half-wave stack, but
		the layer spacings at the right-hand end are 1.4 times the spacings
		at the left-hand end.}
	\label{fig:chirp}
\end{figure}

\begin{figure}
	\includegraphics[width=0.9\columnwidth]{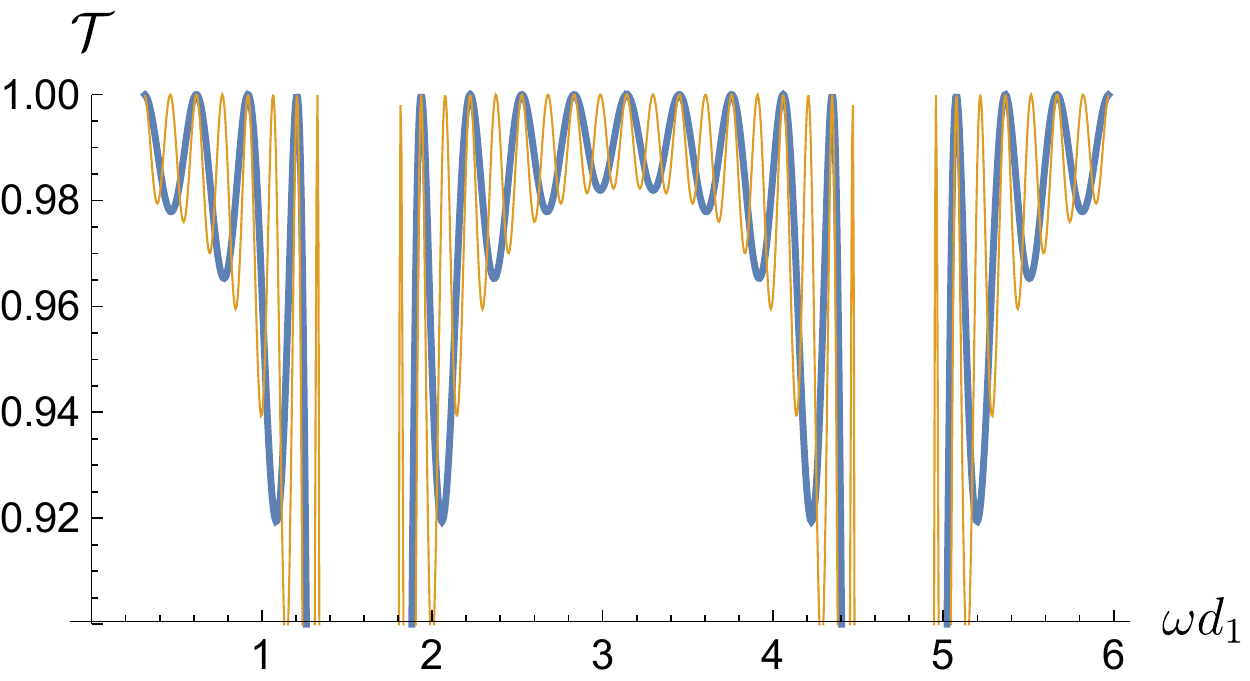}
	\caption{Thick blue (thin orange) curve: transmittance of a 5-period (10-period)
	half-wave stack, with refractive indices
		$n_1 = 1, n_2 = 1.46$ (analogous to alternating gas/silica
		layers --- see Section~\ref{sec:dielectrics}).
	}
	\label{fig:transp}
\end{figure}

\begin{figure}
	\includegraphics[width=0.9\columnwidth]{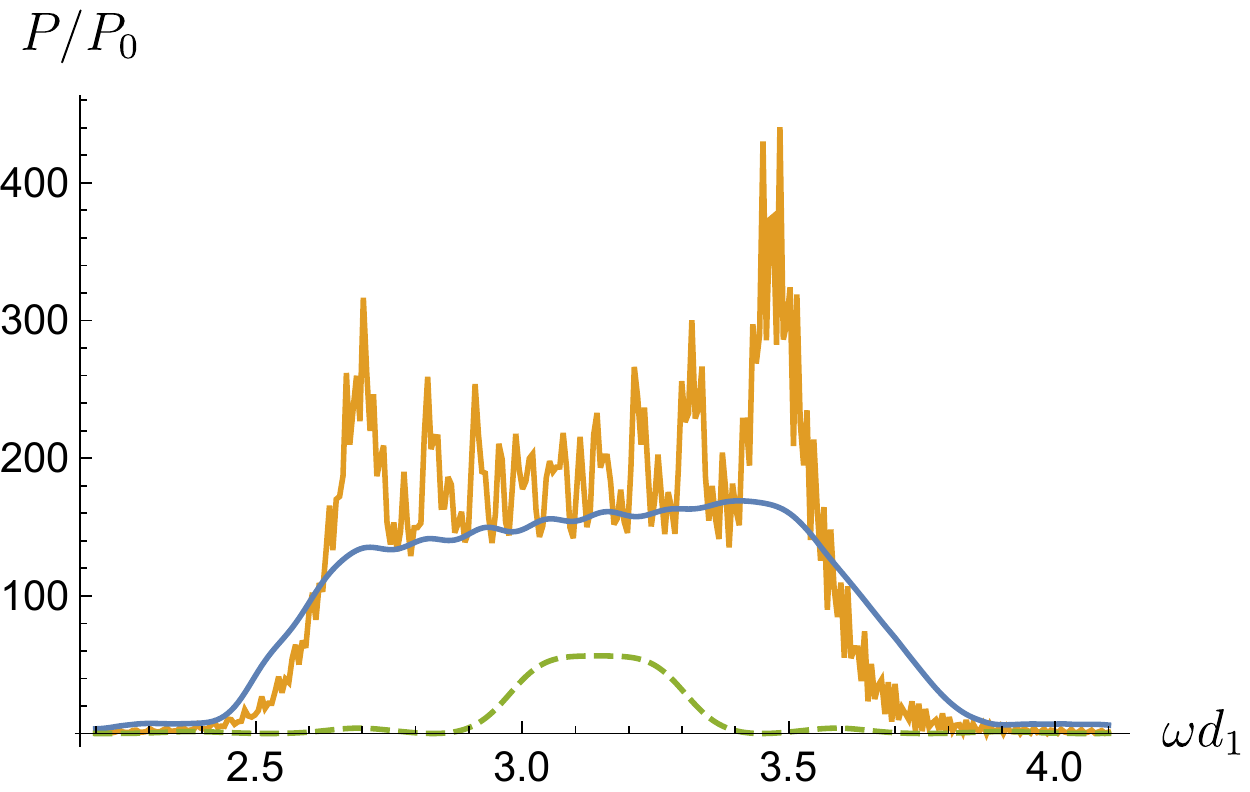}
	\caption{Green dashed curve: power absorbed from spatially uniform DM oscillation
	at a single frequency $\omega$, by a half-wave stack of 10 periods
	with refractive indices $n_1 = 1, n_2 = 1.46$ (analogous to
	alternating gas/silica layers --- see Section~\ref{sec:dielectrics}).
	Blue curve: incoherent sum of powers for 10 different
	10-period half-wave stacks
	at different spacings, covering a fractional frequency range
	of $\sim 30\%$.
	Orange (spiky) curve: power for a 100-period `chirped stack',
	with layer spacings increasing by a factor 1.4 from one end
	to another (similar to Figure~\ref{fig:chirp}).
	While this is spikier than the incoherent
	sum, it still results in an $\OO(1)$ constant converted power across
	the same frequency range.
	The reference power is defined by $P_0 \equiv g^2 B_0^2 \rho_{\rm DM} \omega^{-2}$, for axion DM in a uniform background
	magnetic field $B_0$.}
	\label{fig:layercomb}
\end{figure}

As we increase the number of layers in a half-wave stack,
the frequency range covered decreases as $1/N$. 
It is possible to increase the frequency range covered by a single
stack, by having different parts of it correspond to half-wave
stacks at different frequencies. If this variation is in a direction
parallel to the layers, and is over a scale significantly larger than
the DM coherence length, then the effect is equivalent to running multiple
stacks side-by-side. 

The variation can also be perpendicular to the layers, e.g.\
by gradually changing the layer spacings and/or refractive indices
from layer to layer, as illustrated in Figure~\ref{fig:chirp}.
We refer to this configuration as a `chirped stack'.
It is important that, over the scale of a few layers,
the stack is close to half-wave, since other configurations
are not transparent to nearby frequencies, and result in very spiky frequency
profiles.
Figure~\ref{fig:transp} plots the transmittance of illustrative
half-wave stacks, as a function of frequency, showing how they can be close
to transparent over an order-1 range of frequencies around their central
frequency. In order that the emission from the layers
at one end of a stack is mostly transmitted through the layers at the far
end, the decrease in phase depths should be $\lesssim 30\%$.
The interference effects arising from imperfect transmittance
result in a spiky frequency profile, as shown
in Figure~\ref{fig:layercomb}, but for a wide range of parameters,
the troughs are not large enough to be problematic.

The simplest ways of varying the layer spacings or indices,
such as a smooth variation (as per Figure~\ref{fig:chirp}),
or stacking different half-wave stacks on top of each other,
all result in similar frequency profiles. It seems likely that
this level of spikiness is inevitable, and cannot be ameliorated
by clever choices of layer spacings. Another important point
is that, unlike a half-wave stack, a chirped stack should not
be placed on top of a mirror. Doing so results in very deep
troughs in the frequency profile (effectively, from a
part of the stack interfering with its reflection). 
If the mirror is placed far enough away from the stack that
the light travel time is longer than the DM coherence time, then
such cancellations can be avoided --- however, this corresponds
to $\gtrsim 10^6$ wavelengths, which would most likely be inconvenient 
on laboratory scales.


\section{Experimental Setup}
\label{sec:setup}

\begin{table*}[t]
\centering
\bgroup
\def\arraystretch{1.5}
\begin{tabular}{l | l | l | l}
 & \multicolumn{1}{c|}{Pathfinder} &\multicolumn{1}{c|}{Phase I} & \multicolumn{1}{|c}{Phase II} \\
\hline
Signal & Dark Photon & Dark Photon \& Axion & Dark Photon \& Axion \\ 
\hline
\multirow{2}{*}{Range ($m_{\rm DM} \,\& \,\lambda_{\rm Compton}$)} & $(1\, {\rm eV} , 10\, {\rm eV})$ & $ (50\, {\rm meV} , 10\, {\rm eV} )$ & $(50\, {\rm meV} , 10 \,{\rm eV}) $ \\
& $( 0.1 \,{\rm \mu m} , 1 \,{\rm \mu m})$ & $( 0.1 \,{\rm \mu m} , 20 \,{\rm \mu m})$ & $(0.1 \,{\rm \mu m} , 20 \,{\rm \mu m})$\\
\hline
Area ($A $) &$(10\, {\rm cm})^2$  & $(10\, {\rm cm})^2$ &  $(30\, {\rm cm})^2$ \\
Number of periods ($N$) &   $ 30$ &  $100$ & $1000$  \\
Temperature ($T_{\rm layer}$) & $ 200 \,{\rm K} (300 \,{\rm K})$ & $4 \,{\rm K}$ &  $4 \,{\rm K}$   \\
Thickness ($d \sim N\lambda$) & $(\sim 3 \,{\rm \mu m} , \sim 30 \,{\rm \mu m})$ & $(\sim 10 \,{\rm \mu m} ,\sim 2 \,{\rm mm})$ & $(\sim 100 \,{\rm \mu m} , \sim 20 \,{\rm mm})$  \\
Stacks per $e$-fold & $150$ & $400$ & $4000$  \\
\hline
Detector Dark Count ($\Gamma_{\rm DCR}$) &   $ \, {\rm mHz}$ (e.g. CCD) &  $ 10^{-5}\, {\rm Hz}$ (e.g. TES)&  $ 10^{-5}\, {\rm Hz}$ (e.g. TES) \\
Detector Efficiency ($\eta$) &  $0.1$ &  $ 0.9$ &  $0.9$  \\
Temperature ($T_{\rm detector}$) & $ 200 \,{\rm K}$ & $100 \,{\rm mK}$ &  $100 \,{\rm mK}$   \\
\hline
 Magnetic Field (Axion) & N/A & $10 \,{\rm T}$ &  $10 \,{\rm T}$\\
 \hline
\end{tabular}
\caption{Summary of nominal experimental parameters for the different phases
	of the experiment. As discussed in Section~\ref{sec:tol}, the
 fractional variation in layer thicknesses should be $\lesssim
 1/(\sqrt{2 N})$ for an $N$-period stack. The temperature of the layers
 for the pathfinder phase of the experiment can be either $\sim 200 \,{\rm K}$, which matches the operational temperature of
	the PIXIS CCD photon detector (Section~\ref{sec:photondet}), or as high as room temperature. The number of (mirror-backed) half-wave stacks needed to 
	provide smooth coverage over an $e$-fold in DM mass range is shown.
	Fewer stacks with broader frequency coverage could be used, at the cost
	of lower sensitivity; similarly, multiple stacks could be run
	simultaneously to reduce total integration time.
	}
	\label{tab:tech}
\egroup
\end{table*}

\begin{figure*}[t]
	\includegraphics[width=.9\linewidth]{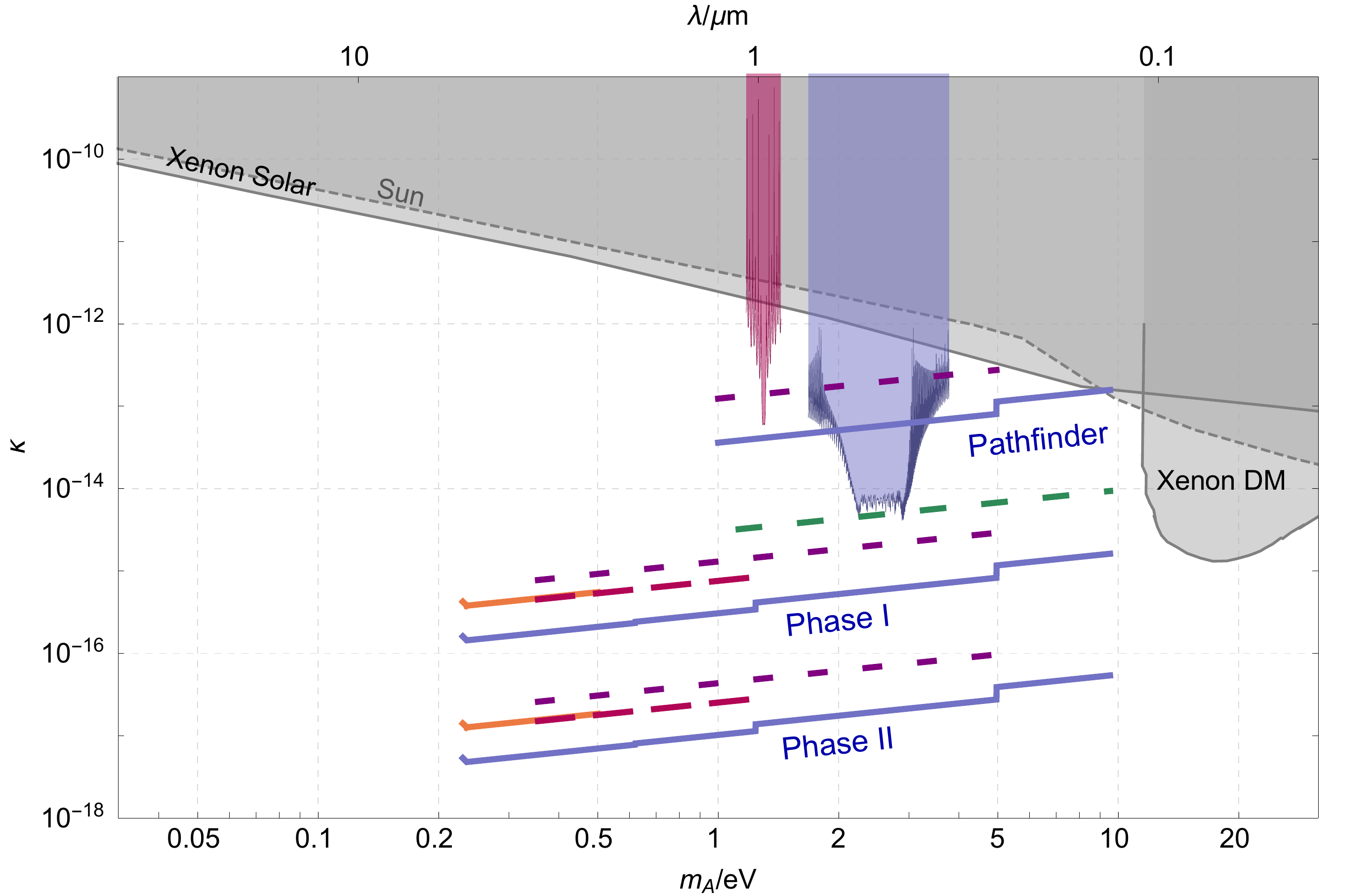}
 \caption{Sensitivity to dark photon dark matter, in terms
 of the kinetic mixing parameter $\kappa$. The different experimental
 configurations are described in Table~\ref{tab:tech}. 
	The reach is shown for a combination of half-wave stacks
	at different spacings, enabling smooth coverage of the mass range
	(Section~\ref{sec:setup}), and is a factor $\sim 2$ deeper than the
	peak reach for a single half-wave stack.
	We assume an integration time of $10^6 \second$ for each stack.
	We also show the sensitivity curves for two example stacks,
	consisting of alternating air / ${\rm Si}_3{\rm N}_4$ layers:
	a $N=30$ half-wave stack (red) with Pathfinder parameters (in particular,
	a detector DCR of $10^{-3} \Hz$), and a
	wider-band $N=100$ `chirped' stack (blue), as per Figure~\ref{fig:layercomb},
	with Phase I parameters.
	The different
 colors/styles of curves correspond to different alternating dielectric
 pairs:
  Ge/NaCl (solid, orange), $\mathrm{SiO}_2$/GaAs (long-dashed, red),
$\mathrm{Si}_3\mathrm{N}_4$/$\mathrm{SiO}_2$ (short-dashed, purple),
doped $\mathrm{SiO}_2$ (dot-dashed, green). The solid
blue lines correspond to alternating air/dielectric structures,
	for (from low to high frequencies) Ge, Si,
$\mathrm{Si}_3\mathrm{N}_4$, $\mathrm{SiO}_2$.
	Different assumptions about the DM velocity distribution will
	only have a small effect on sensitivity (see Section~\ref{sec:sens}).
	The dotted reach at low energy is an estimate of single-photon detector sensitivity in the IR (Section~\ref{sec:photondet}).
	Gray regions indicate current constraints from direct detection
experiments \cite{An:2013yua,An:2014twa} and astrophysical measurements
\cite{Gondolo:2008dd, Vinyoles:2015aba}. 	}
	\label{fig:dpsensitivity}
\end{figure*}

\begin{figure*}[t]
	\includegraphics[width=.9\linewidth]{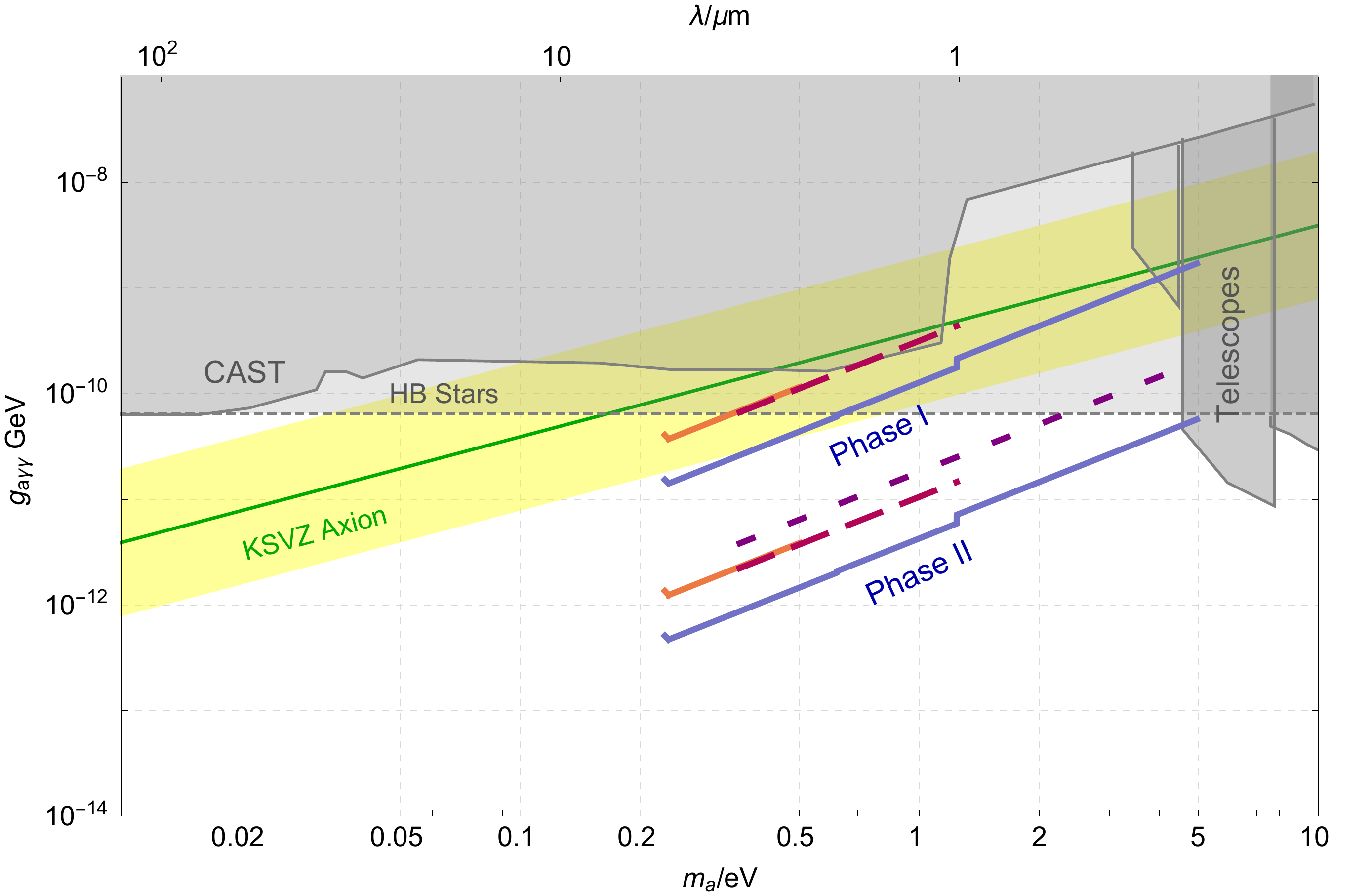}

 \caption{Sensitivity to axion dark matter, in terms
	of its coupling to photons $g_{a \gamma \gamma}$. The different experimental
 configurations are described in Table~\ref{tab:tech}. The reach is
 shown for a series of dielectric stacks at different spacings, enabling
 smooth coverage of the mass range (Section~\ref{sec:setup}). We assume
 an integration time of $10^6 \second$ for each stack. The different
 colors/styles of curves correspond to different alternating dielectric
 pairs:
  Ge/NaCl (solid, orange), $\mathrm{SiO}_2$/GaAs (long-dashed, red),
$\mathrm{Si}_3\mathrm{N}_4$/$\mathrm{SiO}_2$ (short-dashed, purple),
doped $\mathrm{SiO}_2$ (dot-dashed, green). The solid
blue lines correspond to alternating air/dielectric structures,	for (from low to high frequencies) Ge, Si, $\mathrm{Si}_3\mathrm{N}_4$.	The dotted reach at low energy is an estimate of single-photon detector sensitivity in the IR (Section~\ref{sec:photondet}).	Gray regions indicate current constraints from CAST \cite{Anastassopoulos:2017ftl}, stellar cooling \cite{Schlattl:1998fz,Vinyoles:2015aba, Schlattl:1998fz, Ayala:2014pea}, and axion to photon decays \cite{Grin:2006aw}. 
 The Kim–Shifman–Vainshtein–Zakharov (KSVZ) axion line is shown as a guideline; SN 1987A constraints on the nuclear coupling of the QCD axion limit the mass to be $m_a\lesssim 60\meV$ \cite{Chang:2018rso, Patrignani:2016xqp},	assuming that its derivative couplings to	nucleons are not suppressed.
	}
	\label{fig:sensitivity}
\end{figure*}


Figure~\ref{fig:setup2d} shows a sketch of the experimental setup
outlined above; a stack of dielectric layers is placed in a shielded
volume, and then photons emitted normal to the layers
are focused onto a small detector.
A dielectric stack in free space will emit equally in both directions; to facilitate detection,
one end is terminated with a mirror.
For dark photon DM, this setup is sufficient, while for an axion-photon
coupling, a large background $B$ field parallel to the layers would be
introduced.
In this Section, we develop this outline into a more detailed illustration
of how such an experiment might be implemented.

To recap the most important properties of such a setup,
the emitted photon power for $N$ periods of alternating
dielectrics, with refractive indices $n_1, n_2$, at the frequency $m_0$
for which they form a half-wave stack (see Figure~\ref{fig:emodes}),
\begin{equation}
	m_0 = \frac{\pi}{n_1 d_1} = \frac{\pi}{ n_2 d_2},
	\label{eq:m0}
\end{equation}
 is
\begin{align}
	\langle P_{\rm abs} \rangle 
	&\simeq g^2 B_0^2 \frac{\rho_{\rm DM}}{m^2} Q A N 
	\left(\frac{1}{n_1} + \frac{1}{n_2}\right) \left(\frac{1}{n_2} - \frac{1}{n_1}\right)^2\nonumber
	\\
	&= 
	g^2 B_0^2 \frac{\rho_{\rm DM}}{m} \frac{Q V}{\pi}
	\left(\frac{1}{n_2} - \frac{1}{n_1}\right)^2,
\end{align}
where $Q$ is the quality factor of the mode in the layers,
and $V$ is the volume of the stack.
This expression is valid as long as the depth of the stack
is smaller than a DM coherence length, corresponding to 
$N \lesssim v^{-1} \sim 10^3$ (See Section~\ref{sec:otherconf}).
The quality factor depends on how the layers are terminated, but is $\propto N$.
For example, the setup drawn in Figure~\ref{fig:setup2d}, with a mirror
on one side and air on the other, has $Q \simeq 4 N (1/n_1 + 1/n_2)$.
For a dark photon, the expression is
\begin{equation}
	\langle P_{\rm abs} \rangle 
	= \frac{2}{3} \kappa^2 m \rho_{\rm DM} \frac{Q V}{\pi}
	\left(\frac{1}{n_2} - \frac{1}{n_1}\right)^2,
	\label{eq:powerdarkphoton}
\end{equation}
after averaging over many DM coherence times.
These expressions apply over a fractional frequency range $\sim 1/Q$;
the converted power as a function of frequency is non-Lorentzian
(see Figure~\ref{fig:pcomp}), and has peak power 
at frequencies $\sim 1/Q$ away from the center frequency $m_0$.

The mass-averaged conversion power from a 
half-wave stack, over the DM mass range $0 < m < 2 m_0$, is
\begin{equation}
	P_{\rm av} \simeq 
	\begin{cases} 
		2 g^2 B_0^2 \frac{\rho_{\rm DM}}{m_0}\frac{ V }{\pi}
		\left(\frac{1}{n_2} - \frac{1}{n_1}\right)^2 & \mbox{(axion)} \\
		\frac{4}{3} \kappa^2 m_0 \rho_{\rm DM} \frac{V}{\pi}
		\left(\frac{1}{n_2} - \frac{1}{n_1}\right)^2 &\mbox{(DP)}
	\end{cases},
	\label{eq:pav}
\end{equation}
where DP stands for dark photon.
If we construct $N_s$ different half-wave stacks, spaced
to cover a frequency range $\Delta \omega$ with central frequency $m_0$,
then the converted power averaged over the $\Delta \omega$ frequency range
is 
$P_{\rm sum} \simeq N_s \frac{2 m_0}{\Delta \omega} P_{\rm av}$
(if all of the stacks have the same number of layers $N$, and $\Delta \omega \ll \omega$).
Since $N_s \simeq  Q \Delta \omega / m_0$ stacks are required
to obtain a smooth frequency profile, $P_{\rm sum} \simeq 2 Q P_{\rm av}$.
For mirror-backed stacks, with $Q \simeq 4 N (1/n_1 + 1/n_2)$, this gives
the signal power in our experiment,
\begin{equation}
	P_{\rm sum}
	\simeq 
	\begin{cases} 
		8 g^2 B_0^2 \frac{\rho_{\rm DM}}{m_0^2} A N^2
		\frac{(n_1^2 - n_2^2)^2}{n_1^4 n_2^4} &\mbox{(axion)} \\
		\frac{16}{3} \kappa^2 \rho_{\rm DM} A N^2
		\frac{(n_1^2 - n_2^2)^2}{n_1^4 n_2^4} & \mbox{(DP)}
	\end{cases}.
	\label{eq:psum}
\end{equation}
These expressions can be used to estimate the sensitivity for a set of half-wave stacks covering some DM mass range, Figs.~\ref{fig:dpsensitivity} and \ref{fig:sensitivity}.

In the following, we consider three illustrative stages for the
experiment (see Table~\ref{tab:tech}). The first, a `pathfinder', would
be the simplest that could still probe new parameter space; it would aim
to detect dark photon DM, could be run at room temperature, and would
use readily-available detectors and layer fabrication methods. Phase~I
would aim to explore significant new dark photon parameter space, and
start gaining sensitivity to new axion parameter space. This would
require cooling the target to cryogenic temperatures, using cryogenic
detectors, and using $\OO(100)$ high-contrast layers, as well as
operating with a large background $B$ field for the axion search. Phase
II would aim to cover significant new axion and further dark photon
parameter space, using a larger volume of layered material.


\subsection{Dielectric materials}
\label{sec:dielectrics}

\begin{table}[t]
\centering
\begin{tabular}{l | l | l | l | l|l}
  & $n$ & Wavelength & DM mass & CTE & Radioactive\\ 
    &  & range (${\rm \mu m}$) & range (eV) &  ($10^{-6}/{\rm K}$) &\\
    \hline
${\rm SiO_2}$ & 1.46 & (0.13, 3.5) & (0.35, 9.5) & 0.55 &   \\
GaAs & 3.8 & (1, 15) & (0.08, 1.2) & 5.7 &     \\
Ge & 4.0 & (2, 17)  & (0.07,0.62) & 6.1	 &   \\
NaCl & 1.49 & (0.2, 20) & (0.06 , 6.2) & 44 & ${\rm {}^{36} Cl}$~\cite{strauch2014isotope,Patrignani:2016xqp} \\
${\rm Si_3 N_4}$ & 2.00 &(0.25, 8) & (0.15, 5)& 3.3 &  \\
Si & 3.42 & (1.1, 9) & (0.12, 1.0)& 2.55	 &  \\
${\rm MgF_2}$ &1.41 & (0.12, 9.5) & (0.13 ,10) &13.7	 &  \\
${\rm CaF_2}$ &1.43 & (0.15, 9) & (0.14, 8.3)& 18.85	&  \\
${\rm ZnSe}$ & 2.4 & (0.55, 20) & (0.06, 2.3) & 7.1	 &  ${\rm {}^{79}Se}$~\cite{Patrignani:2016xqp}  \\
${\rm GaN}$ & 2.3 & (0.37, 13.6) & (0.09, 3.4) & 3.17	 &   \\
\hline
\end{tabular}
	\caption{A list of materials that can be used to construct the layers for different dark photon and axion masses and their main optical and thermal properties~\cite{HHLi1980,EdmundO,rii,PhysRevLett.102.226401,quartztemp,Ligentec,ISPO,SOMA1982889}. The refractive index ($n$) is the value at vacuum-wavelength $\sim 1.5 \,{\rm \mu m}$, at room temperature. The pathfinder phase of our experiment will be operated at $\sim 200-300\,{\rm K}$, while Phase I and Phase II of our experiment will be operated at liquid helium temperatures. Refractive indices at those temperatures will be needed for manufacturing purposes. Coefficients of thermal expansion (CTE) at room temperature are also listed. A few of the elements have radioactive isotopes that can cause additional backgrounds in our experiment, and therefore require additional purification procedures.}\label{tab:material}
\end{table}
As discussed above, the proposed experiment requires periodic structures with a significant refractive index contrast; for lower frequencies, these need
to operate at cryogenic temperatures.
In this Section, we summarize the requirements on material properties necessary for our setup. 

The layers should be transparent at the relevant frequencies. to avoid losses; transparency windows, typically extending up to the bandgap energy, 
are listed in table~\ref{tab:material}. For example, transmittance in excess of $99.9\%$ at wavelengths above 250 nm is demonstrated for silica with $\mm$ thickness~\cite{JEOS:RP13010}. 

An interesting phenomenological property of crystal dielectrics is that the refractive index decreases with increasing bandgap energy $\omega_0$:  $n^4\omega_0 \sim 100\eV$ \cite{Moss2,MossReview}. Thus with common dielectrics it is easier to achieve high sensitivity at lower frequencies.

Another requirement comes from the fact that, in order to suppress
thermal backgrounds at frequencies $\lesssim \eV$, it will be necessary
to cool the dielectric stack to cryogenic temperatures (see
Section~\ref{sec:bb}). For the layers to be stable under this
temperature change, the materials should ideally have similar
thermal expansion properties. One possibility would be to use
the same host material --- for example, the widely used silica
(${\rm SiO_2}$) --- with index-raising (e.g. Germania (${\rm
GeO_2}$)) and index-lowering (e.g. Boron trioxide (${\rm B_2
O_3}$) and Fluorine (${\rm F}$)) dopants for the alternating
layers~\cite{RevModPhys.78.1135,BUTOV2002301,HERMANN19891083}. A
refractive index contrast of up to $10\%$ can be achieved without
significantly altering the mechanical and thermal properties of the
material~\cite{Bachmann1988,Sinha}. This procedure has been studied
extensively e.g.\ to improve the performance of optical fibers at low
temperatures~\cite{Hashimoto}, which rely on a significant core-vs-cladding
refractive index contrast created by adding dopants to the host material
in the fiber core~\cite{ziemann2008pof,m1992optical,RevModPhys.78.1135}.

For an improved version of the experiment, materials with $\OO(1)$
different refractive indices should be employed: maximum dark matter
to photon conversion is achieved when the dielectric materials have
indices of refraction $n_1=1,\, n_2 \gg 1$ (eq.~\eqref{eq:pav}). Thus, the
highest power results when alternating air or vacuum with materials
with high refractive index such as silicon (Si, $n \sim 3.4$). Layering
alternating dielectrics may be more mechanically robust, in which
case pairs such as silica ($\mathrm{SiO}_2,\,n \sim1.46$) and gallium
arsenide (GaAs, $n\sim 3.8$), commonly used in the semiconductor
industry~\cite{Sinha}, can achieve $\sim 20\%$ of maximum power. It
will be necessary to demonstrate that at least tens of alternating layers can be
constructed and withstand the thermal stresses of cooling to cryogenic
temperatures~\cite{SOMA1982889,WANG20075500}.

The relevant technologies to create layered structures vary
depending on the scale of the desired spatial periodicity, set by
the dark matter mass. For DM in the $0.1 - 10 \eV$ mass range,
corresponding to layer thicknesses $\sim 0.1 - 10 \um$, possible
production processes include chemical vapor deposition (CVD), physical
vapor deposition (PVD), spin coating, epitaxy, and sputtering
(for a review, see~\cite{seshan2012handbook}). These methods are
well-established for such length scales and are employed in producing
optics such as mirror and lens coatings, and semiconductor diode lasers
(surface-emitting laser and the vertical-cavity surface-emitting laser
(VCSEL))~\cite{Iga2000,PrincetonO,Zhang:14}, with tens of layers in
the optical wavelength range commercially available. Acid-etching
can potentially be used to achieve alternating dielectric/air
structures~\cite{Russell358,Tonucci783}.

As discussed in Section~\ref{sec:tol}, the structure should be
close to flat and periodic. Each stack, once built, can be
tested with an broadband laser beam, for example a Ti:Sapphire
laser \cite{Moulton:86}. By measuring the transmitted and reflected
wave from the stack, the resonant frequency as well as the normal
direction of the layers can be measured to very high precision
\cite{photonicbook,Noda:16,johnson2001photonic}.

Our calculations throughout have neglected dispersion, by
assuming that the dielectric materials have a constant refractive index
as a function of frequency.
For the fairly small fractional frequency ranges covered by each
stack, this is almost always a good approximation, since refractive
index changes over $e$-fold frequency ranges within the bandgap are $\lesssim 5\%$. 


\subsection{Photon detection}
\label{sec:photondet}

As discussed in Section~\ref{sec:sens}, the existing constraints on
axion or dark photon couplings mean that, even if they make up all of
the dark matter, the photon conversion rate from a reasonably-sized
target will be small. Accordingly, we will require a sensitive photon
detector with a sufficiently low energy threshold, high photon efficiency,
and low noise.

From Section~\ref{sec:halfwave}, the DM momentum spread means that
converted photons are emitted in a narrow cone (opening angle $\sim
10^{-3}$) around the normal to the layers. Consequently, the photons
from a stack of cross-sectional area $A$ can be optically focused
down to an area $\sim 10^{-6} A$ (for stack radii $\gtrsim \cm$,
this is larger than a square wavelength, for the DM mass range we are
considering). For a stack of area $(10 \cm)^2$, the detector
must have area $\gtrsim (100 \micron)^2$ to intercept $\OO(1)$
of the signal photons. The required area could be decreased by various
techniques, such as using a high-refractive-index concentrator
on top of the detector, or having the signal photons bounce multiple
times within a cavity.

The most commonly used photon detectors in our frequency range are
charge-coupled devices (CCDs), which are found in a wide range of astronomical and
laboratory applications. A close analogue to our low-signal-flux,
long-integration-time setting is the ALPS `light shining through
wall' experiment~\cite{Bahre:2013ywa}, which looks for very rare
photon-axion-photon conversion events at optical frequencies. The PIXES
CCD camera~\cite{PIXISCCD}, used in the ALPS experiment and planned as a backup for the ALPS-II upgrade, operates
at $\gtrsim -70 {}^{\circ} {\rm C}$ and has $1024 \times 1024$ pixels
with per-pixel area of $ (13 {\rm \mu m})^2$, detection efficiency of
$\eta\sim 10 \%$, and dark count rate $\Gamma_{\rm DCR} \sim \,\rm{mHz}$
for wavelengths shorter than $\sim \mum$. Accordingly, we adopt similar
parameters for the pathfinder stage.

Detectors with similar frequency coverage ($ m_{\rm DM} \gtrsim
1.1\,{\rm eV} $), better efficiency ($\gtrsim 20 \%$) and lower dark count
rate in cryogenic environments ($0.1 \,{\rm mHz}$ per pixel)~\cite{private}, are
being developed for dark matter direct detection experiments based on
liquid xenon. The per-pixel area of this detector is $50 \mum \times 50
\mum$, and arrays of $60\times 60$ pixels have been demonstrated \cite{Arneodo2018}. With
optimization, these detectors can be ideal for transitioning between
the pathfinder and phase I of our experiment. 

To reach our phase I and phase II sensitivities, dark count rates of
$\lesssim 10^{-5} \Hz$ are required. These rates have been demonstrated
for multiple detector technologies, including Transition Edge Sensors
(TES)~\cite{Dreyling-Eschweiler:2014mxa,Dreyling-Eschweiler:2015pja,Cabrera1998,Karasik:2012rb,Lita:08,Bastidon:2015aha} (for a review, see~\cite{Irwin2005}), Microwave Kinetic Inductance
Detectors (MKIDs)\cite{Mazin,DayLeduc,GaoMazin}, and nanowires
\cite{Rosfjord:06}. The efficiency of these detectors can be quite
high: for example, a TES can be coupled to photon modes using
specially designed coatings in a narrow range of frequencies, reaching
efficiencies of $\eta\gtrsim 95\%$, with demonstrated $98 \% - 99\%$
detection efficiency for wavelengths between $0.6$ and $2\mum$
\cite{Lita:08,Karasik:2012rb}.

Achieving such low dark count rates generally requires small detectors.
For example, a TES has an exponentially suppressed dark count rate
above its energy resolution, but this energy resolution increases with the size
of the TES;
thus, to keep the dark count rate low, it is crucial that we
are able to focus the signal to a small area, of order tens of microns on a side ~\cite{Cabrera1998,Irwin2005,Lita:08}.
One possibility to achieve a larger total detector area
is to multiplex multiple TES pixels; arrays of more than 200 pixels have
been demonstrated~\cite{Irwin2005}. 

For most of the energy range we cover, we assume that the DCR
and other backgrounds (see Section~\ref{sec:bkg}) can be controlled
to below $10^{-3} \Hz$ for
the pathfinder phase, and below $10^{-5} \Hz$ for phases I and II. 
At DM masses below $\sim 0.2 \eV$, the energy of a signal photon
is close to the currently achievable detector energy resolution.
For the low-frequency regions of Figures~\ref{fig:dpsensitivity} and~\ref{fig:sensitivity}
(shown in dotted lines), we
assume a TES-type noise curve with 
$\Gamma_{\rm DCR} \propto 0.1 \Hz \, e^{-\omega^2/2\Delta E^2}$,
with $\Delta E \sim 50 \meV$. This normalization 
gives a reach that approximately matches to the reach of a bolometer with noise equivalent power $\sim 10^{-20}
\mathrm{W}/\sqrt{\Hz}$ (demonstrated at lower energies~\cite{Farrah:2017moj,Bradford2015,Irwin2005}) at around $50 \meV$.
We are not aware of whether such sensitivities have been
demonstrated in the near- to mid-IR regime we are considering.

The signal power calculations above have used the framework of a
classical DM field driving a classical signal photon mode. For the
light bosonic DM production mechanisms discussed in Section V, the
early-universe DM abundance is expected to take the form of a large
occupation number coherent state, corresponding to a classical-like
oscillation of the field, with a definite oscillation phase. Since then,
the evolution of the coherent state has most likely maintained this
coherence. Even though the occupation number of DM modes around Earth
is less or comparable to $1$ for $m \lesssim 20\, \mathrm{eV}$, these
will still be small-amplitude coherent states (like e.g.\ an attenuated
laser mode). The DM then excites the signal mode into a small-amplitude
coherent state, which has mean occupation number given the the classical
power calculation, and Poissonian number statistics. Over timescales
longer than the DM coherence time, variation of the DM field amplitude
can lead to super-Poissonian photon detection fluctuations. If the DM
is significantly spatially clumped, then these fluctuations can be very
large; however, the most common assumption is that most of the galactic
DM is smoothly distributed, in which case such fluctuations will average
out over many coherence times.

If --- due to some unknown mechanism --- the DM field around Earth were
in a very different quantum state, then we would still expect almost
the same Poissonian statistics of detected signal photons. This is true
whenever the probability of converting a DM mode to a photon is small.
Hence, our sensitivity calculations should apply very generally. These
considerations justify treating the DM as a classical-like background
oscillation, as done in many light bosonic DM detection methods.
Potential experimental differences between different DM quantum states
can arise if e.g.\ we employ a phase sensitive amplification method,
rather than pure photon counting. 


\subsection{Scanning}\label{sec:scan}

The simplest scanning mechanism would be if it were possible
to change the refractive indices of the layers by an external
perturbation. If the refractive indices could be changed by $\OO(1)$,
this would enable a single stack to cover an entire decade of frequency
range. However, the optical materials of the types we have been
considering typically have very small refractive index changes in
response to reasonable external perturbations. For example, the rate of
change of refractive index with temperature for silica, at $\sim \mum$
wavelengths, is $\lesssim 10^{-5} {\rm \, K}^{-1}$~\cite{quartztemp}. Especially
at the lower end of our frequency range, where the layers' temperature
needs to be low to suppress blackbody radiation, this does not allow
significant scanning. Similarly, the refractive index changes due to
applied
electric fields (Kerr/Pockels effect), magnetic fields (Verdet effect) or
strain (photoelastic/piezooptic effect) are generally too small for our purposes.

While birefringent materials can have different refractive indices for
different propagation directions, there are only two distinct
polarizations corresponding to a given propagation direction.
Since the direction of photon emission is set (to within $\sim 10^{-3}$)
by the layered structure of the material, changes that do not affect
the layered structure itself (e.g. rotating the material relative to the $B$ field) will not result in significant scanning.

The opposite question, of whether we can tunably \emph{narrow} the bandwidth
of an existing stack, is also of interest. When backgrounds
are important, narrowing the bandwidth improves the signal to noise,
and it would also allow us to home in on a tentative signal.
The simplest way to reduce the bandwidth is to increase
the quality factor by placing e.g.\ a half-wave stack inside a `cavity'.
For example, if we place a quarter-wave stack of $M$ periods
above a half-wave stack, this increases the $Q$ factor by $G = C^{M}$,
up to manufacture and alignment accuracy
(where $C$ depends on the dielectric contrast of the quarter-wave stack).
By changing the separation between the quarter-wave and half-wave
stacks over a distance $\sim N/G$ wavelengths, we can scan this narrowed bandwidth
across the entire original bandwidth of the half-wave stack.
This procedure would demand improved tolerances so as not to smear
out the narrowed peaks -- in particular, the separation between the plates should be the same across the whole area, to within
$\sim N / G^2$ of a wavelength -- and accurate positioning of the quarter-wave
stack to within $\sim N/G^2$ wavelengths.


\subsection{Environmental backgrounds}\label{sec:bkg}

Since the flux of signal photons in our experiments would be very
weak, it is important to be able to discriminate these from backgrounds.
In this Section, we will discuss the backgrounds from radioactivity,
cosmic rays, and blackbody radiation, and the requirements these impose
on our experimental design.

\subsubsection{Blackbody}
\label{sec:bb}

If the detector's field of view is at temperature
$T$, the rate at which thermal photons within a small energy
range $\Delta \omega$ of $\omega$ hit the detector is
\begin{equation}
	\Gamma_{\rm BB} \sim \frac{\Delta \omega \, \omega^2}{4 \pi^2} A_{\rm det} e^{- \omega/T},
\end{equation}
for $\omega \gg T$, where $A_{\rm det}$ is the area of the detector. For
the pathfinder, a room-temperature
field of view ($T \sim 300 \K$) 
gives small ($\lesssim {\rm \, mHz})$ dark-count rate for $\omega \gtrsim \eV$. Since
$\omega \gtrsim 40 T$, we are well into the blackbody tail, and
the minimum frequency giving the desired dark count rate depends only logarithmically on detector size.
For phases I and II, we are interested in
photon energies down to $\sim 50 \meV \simeq 600 \K$, 
and similarly require $T \lesssim \omega / 40 \sim 15 \K$.
Since we are focusing the signal photons from the layers
onto the detector, we require that at least the dielectric layers,
and the surrounding shielding, are cooled to these temperatures.
It would likely be practical to cool
the target volume to $4 \K$ using liquid helium.

The photodetectors we require for phases I and II are generally
cryogenic, and need to be operated at temperatures $\ll 15 \K$;
for examples, low-noise TESs are operated at $\sim 100 {\rm \, mK}$.
To help achieve this, 
a cold filter could be placed between the 
layers and the detector, which is transparent to photons
at the signal frequency, but blocks the lower-frequency thermal
radiation.

\subsubsection{Cosmic Rays \& Radioactivity}

In addition to blackbody photons, there will also be
less frequent but more energetic background events.
One source of these is cosmic rays.
The sea-level cosmic ray flux is dominated by muons, with a rate
of $\sim 1/(10 \cm)^2/\sec $. These deposit $\sim 100 \keV / \mm$
as they travel through typical materials~\cite{grieder2001cosmic}.
Radioactive decays in or near the experiment constitute another
background. Laboratory materials will contain some pre-existing (or
cosmogenic) level of radioactive isotopes; when these decay, they can
produce particle showers in the experiment. 

The fact that all of the
signal photons are focused onto a small detector, compared to a more uniform
flux of shower particles (from decays or cosmics), improves our signal
to background ratio. In addition, the fact that showers
generally consist of many particles, while signal photons
arrive one at a time, provides a discrimination strategy.
Multiple detectors, either in the form of a pixel array or separate 
detectors, will generally register many simultaneous counts
for a shower event, allowing that time interval to be vetoed. 
As long as $\OO(1)$ of the observation time is not vetoed, these
detectors can have higher dark count rates than the one employed for the
signal photons.

If such veto schemes are not sufficient, then it may be possible
to reduce the background rates directly. For cosmic rays, this could
be accomplished by running the experiment deep underground.
For radioactive decays, it may be possible to
to fabricate the experimental setup with radiologically pure materials.
Taking the example of chlorine, 
the radioactive isotope ${}^{36}\mathrm{Cl}$ has
natural fractional abundance $7\times 10^{-13}$, and half life
$3\times 10^5\,\mathrm{yrs}$~\cite{ensdf}. This gives
rise to an event rate of $\sim 10^{-3} \Hz$ from a gram of chlorine.
A much lower ${}^{36}\mathrm{Cl/Cl}$ ratio of $\sim 10^{-15}$ can
be found in old ground water~\cite{strauch2014isotope}, which would
reduce the background event rate below $\sim 10^{-5} \,{\rm Hz}$.
Other isotopes may be more problematic; for example,
Potassium-40 (${}^{40}{\rm K}$)
has a half life
of $\sim 10^9 \,\mathrm{yrs}$ and natural fractional abundance
of $10^{-4}$~\cite{ensdf}, producing background events at
$\sim 10 \Hz / {\rm g}$.
Table~\ref{tab:material} provides details for isotopes relevant
to the dielectrics listed for layer construction.
For a discussion of other
elements relevant to the rest of the experimental setup, see
\cite{Arvanitaki:2017nhi,Arpesella:2001iz,Iga2000}.

Radioactivity can also be cosmic-ray induced.
For example, high energy
electrons in cosmic rays can produce radioactive ${}^{14} \mathrm{C}$
from stable ${}^{14} \mathrm{N}$. If this must be avoided,
the experiment may have to be run underground.
Similarly, some layer synthesis processes employing high
energy electrons or ions could result in accidental 
production of radioactive elements
(e.g. ${}^{14} \mathrm{N} +e^-\rightarrow {}^{14} \mathrm{C}$ from
stray electrons in certain types of pulsed laser deposition).

It seems likely that some combination of small detector area, vetoing,
and possibly purified materials and/or an underground laboratory, will allow
us to discriminate signal photons from backgrounds in our setups.
Similar backgrounds will affect other light DM experiments, and as
prototypes of these are tested (e.g.~\cite{Crisler:2018gci}), we will gain more information on their
properties. 


\subsection{Sensitivity}
\label{sec:sens}

Figures~\ref{fig:dpsensitivity} and~\ref{fig:sensitivity} show
the projected reach of our different experimental phases (Table~\ref{tab:tech})
for dark photon and axion DM.
The mass range we consider is a well-motivated target for bosonic dark matter
searches: dark photon dark matter is naturally produced by inflationary
perturbations in the early universe, while for axions, a combination of
inflationary perturbations and decays of non-perturbative defects can
produce dark matter densities (Section~\ref{sec:dmprod}).

For the sensitivity plots, we assume that the dark photon or axion
makes up all of the local DM density, $\rho_{\rm DM}\simeq 0.3\GeV/\cm^3$.
To obtain smooth coverage over an $e$-fold range in DM mass,
we require $\sim Q$ half-wave stacks with quality factor $Q$,
at fractional frequency spacings $\sim 1/Q$.
Equation~\eqref{eq:psum} gives the converted power from the incoherent
sum of these stacks.
The sensitivity curves take an exposure time of $10^6 \second$
for each stack. 
For the 30-layer pathfinder configurations shown
in Figure~\ref{fig:dpsensitivity}, covering an $e$-fold mass range
requires $\sim 150$ stacks; a 1 year total integration
time would require running five stacks simultaneously.
For the 100-layer (1000-layer) configurations in phases I and II,
$\sim 400$ ($\sim 4000$) stacks can cover an $e$-fold reasonably
smoothly. As discussed in Section~\ref{sec:chirp}, stacks at different
frequencies can be run on top of each other, at the expense of a slightly
spikier frequency profile. Figure~\ref{fig:dpsensitivity} also shows an example
of this `chirped stack' configuration, illustrating how
a fractional frequency range of $\sim 30\%$ can be covered by a single configuration.

In Figures~\ref{fig:dpsensitivity} and~\ref{fig:sensitivity}, 
we assume the most optimistic simultaneous-running case,
such that a set of adjacent-in-frequency stacks are aimed
simultaneously at the same detector. This means that there is no
dark count noise penalty from having the signal photons for
a given DM mass coming from multiple
stacks. If adjacent stacks have independent detector noise,
then the coupling sensitivity will be degraded, though only by a factor of  $(\mbox{noise counts})^{1/4}$.

The different curves within each phase of Figures~\ref{fig:dpsensitivity}
and~\ref{fig:sensitivity} represent different material pairs
for the dielectric stacks. Section~\ref{sec:dielectrics} describes how
a range of dielectric materials could be used to achieve close-to-optimal
conversion rates over our entire range of frequencies, 
$\sim 50\meV$ to $\sim 10\eV$ (see Table~\ref{tab:material} for examples).
The lower end of this frequency range is limited by the threshold
energy of low dark count single-photon detectors
(Section~\ref{sec:photondet}; see Section~\ref{sec:discussion}
for a brief discussion of bolometric detectors).
The upper end is limited by the availability of simple dielectrics
with low losses and high refractive indices at high
frequencies, and is already well-constrained by other experiments
and astrophysical observations.

Given the stringent constraints on the axion-photon coupling, Phase I
parameters, with the addition of a $\sim 10$ Tesla magnetic field applied
parallel to the layer surface, are necessary to improve on current
bounds, Fig.~\ref{fig:sensitivity}. Phase II can significantly improve
current sensitivities to axion dark matter, which couples to photons, in the mass
range of $0.1-5\eV$. The reach in coupling scales directly with applied
B field, so large fields are required. However, we do not require a
large volume: one stack can occupy as little as $(10 \cm)^2 \times
10\mum = 0.1 \cm^3$. Large magnetic fields with the above-mentioned volumes have been demonstrated~\cite{Apollinari:2015jfa} for both a resistive DC magnet ($19\,{\rm T}$, bore diameter $19\,{\rm cm}$)~\cite{19TMag} as well as a superconducting magnet ($21.1\,{\rm T}$, bore diameter $10.5\,{\rm cm}$)~\cite{21TMag}.  The volume versus B field magnitude should be optimized to achieve the largest reach; in particular, if larger fields are available with smaller bore diameters, the area of an individual stack can be reduced and multiple stacks of smaller area can be run at the same time. To maximize the B field volume, mirror optics can be used to guide the light out of the magnet prior to focusing.  

The axion-photon couplings that we could probe in Phases I and II are well
below the KSVZ and DFSZ couplings for this mass range,
respectively. However, for a generic QCD axion, the axion-nucleon
couplings in this mass range are already constrained
by the lack of SN1987A energy loss to
axions~\cite{Chang:2018rso}. We briefly discuss the possible extension
of our experimental setups to frequencies $\lesssim 60 \meV$ in
Section~\ref{sec:discussion}.

If a tentative signal were detected, the relevant mass range
could be studied by increasing the $Q$ factor, 
either by placing an existing stack in a `cavity', or by 
creating stacks at the signal frequency with a larger number of layers
(Section~\ref{sec:scan}). Doing so increases the signal strength as well
as giving finer frequency coverage, allowing for excellent
discrimination from background, and, eventually, characterization of the
signal properties.

Figures~\ref{fig:dpsensitivity} and~\ref{fig:sensitivity} 
do not take into account the dark matter velocity distribution
(similarly for the equations at the start of this Section).
As discussed in Section~\ref{sec:multilayer}, the effect
of the DM velocity distribution, which has 
$\delta v \sim 10^{-3}$, is to spread out the converted power 
as a function of DM mass by a fractional mass range $\sim \delta v$,
while preserving the mass-averaged power.
Consequently, for $N \ll 1000$ periods, the effect of the DM
velocity distribution is expected to be small, while for $N \sim 1000$ periods, the effect on the power output from each individual stack can be $\mathcal{O}(1)$ .
For example, the total power output from a 500-period
stack at a DM mass corresponding to its central frequency
is only reduced by a factor
$\gtrsim 0.8$ if the DM velocity distribution is set
to the `Standard Halo Model'~\cite{McCabe:2010zh} (SHM), as compared
to zero velocity spread.\footnote{For
a non-mirror-backed stack, the difference between the power
emitted from each end can be significantly larger,
as calculated in~\cite{Millar:2017eoc,Knirck:2018knd}.}
For the 1000-period stacks assumed for Phase II,
the peak power is reduced by $\gtrsim 0.5$
for the SHM, as compared to $\delta v = 0$.
However, since an experiment to cover a wide DM mass range would make
use of many different stacks, with closely-spaced central
frequencies, a wider DM velocity distribution results
in almost the same number of overall converted photon events, but spread
across different stacks. As discussed above,
the effect of this on coupling sensitivity depends on
whether multiple stacks are aimed at the same detector, in which case the sensitivity is as shown in figures~\ref{fig:dpsensitivity} and~\ref{fig:sensitivity}.
In the worst case of separate detectors, 
a reduction in peak power of $\sim 0.5$
results in a coupling sensitivity only a factor of $\sim 2^{1/4} \simeq 1.2$
worse.


\section{Other Dark Matter Candidates}
\label{sec:othercouplings}

Spin-1 DM candidates, such as the dark photon, have an intrinsic 
polarization direction, so can convert to photons even in a
locally isotropic target material. For spin-0 DM candidates, 
there must be some intrinsic direction to the target to
determine the polarization of the converted photon, or else
the conversion rate is suppressed by at least $v^2$, where
$v$ is the DM velocity.
In the case
of an $a E \cdot B$ coupling, a background $B$ field provides
this direction --- however, for couplings to SM fermions, 
the target material itself must have some directionality to avoid
$v^2$ suppression.\footnote{For axions with couplings to fermions, there
is the additional feature that
the target spins must be polarized to get appreciable coherent conversion.}

Here, we give a brief overview of how dielectric haloscopes,
using different target materials, can absorb some other DM candidates.

\subsection{$B-L$ vector}

Apart from the electromagnetic current, the other conserved current
in the SM is $B-L$ (if neutrinos are Dirac). This means that it could
be consistently gauged, and that a new spin-1 particle coupled to $B-L$
could naturally be light without running into strong constraints from
non-renormalizable couplings. Phenomenologically, in situations
where interactions with nuclei are subdominant (for example,
refraction in a dielectric), a $B-L$ vector with coupling
$\LL \supset g_{B-L} X_\mu J_{B-L}^\mu$ will behave like a dark
photon with $\kappa = g_{B-L}/e$. Consequently, our experiments
will absorb $B-L$ DM in the same way as they would dark photon DM.

A difference from the dark photon case is that the couplings
of a $B-L$ vector to neutrons and neutrinos result in stronger
constraints.
The coupling to neutrons
results in a fifth force between neutral matter~\cite{Adelberger:2003zx},
while a DM abundance with $m_{B-L} \gtrsim 0.1 \eV$ can decay to neutrinos,
resulting in cosmological bounds \cite{Gong:2008gi,Wang:2010ma}.
These constraints mean that even our Phase II experiment could
only just reach new parameter space.


\subsection{Scalar couplings}

New light scalar particles (i.e.\ with couplings to
even-parity SM operators) can arise from UV physics in various ways,
e.g.\ Higgs portal models~\cite{Piazza:2010ye}, dilatons / radions~\cite{Damour:1994zq,Taylor:1988nw,ArkaniHamed:1999dz}, 
or moduli fields~\cite{Dimopoulos:1996kp}. For low-energy interactions,
the important couplings are those to EM via $\phi F F$, and
fermion mass couplings $\phi \bar{f} f$ to the electron and nucleons.

The $\LL \supset g_\alpha
\phi F_{\mu\nu}F^{\mu\nu}$ coupling modifies
the Maxwell equations to
\begin{align}
\nabla \cdot E &= \rho + g_\alpha \nabla\phi\cdot E\nonumber\\
\nabla \times B - \partial_t E &= J +  g_\alpha \left(\nabla\phi\times B - \dot{\phi}E\right).
   \label{eq:scalar}
\end{align}
Compared to an axion-type $a F \tilde{F}$ coupling, the
non-velocity-suppressed $\dot \phi$ term now couples to the background
electric field, rather than the magnetic field.
A feasible magnetic field strength of $10 {\rm \, T}$ corresponds,
in natural units, to an electric field strength of $\sim 30 {\rm MV / cm}$ ---
this is much larger than the electric field that can be applied
to standard materials, 
and even in polar
materials, the volume-averaged electric fields are $\lesssim {\rm MV / cm}$.
As a consequence, our experiments would give
sensitivities 
weaker than current constraints from fifth force
tests and stellar cooling~\cite{Adelberger:2003zx,Raffelt:1996wa}.

The scalar couplings to fermion masses, $\LL \supset g_f \phi \bar{f} f$,
give a non-relativistic interaction Hamiltonian (to first order in $g_f$)
\begin{equation}
	H \supset g_f \phi + \frac{g_f \phi}{2 m_f^2} \left((\vec{p} - q \vec{A})^2 - q \vec{\sigma} 
	\cdot \vec{B} \right),
\end{equation}
where $q$ is the charge of $f$, $m_f$ is its mass, and $p$ and $\sigma$
are its momentum and spin operators.
The first term gives rise to a force in the presence of a $\phi$ gradient,
giving velocity-suppressed absorption rates. The second gives
rise to an oscillating magnetic dipole moment, and to
velocity- and acceleration-dependent forces on the fermion. The coupling to 
electron mass, $\phi \bar{e} e$, and the couplings to nucleon
masses, $\phi \bar{n} n$, are subject to strong constraints
from fifth force tests and stellar cooling~\cite{Hardy:2016kme}.
These mean that the velocity-suppressed absorption rate will
not probe new parameter space. 
The other terms could lead to coherent absorption
in directional target materials, but we leave such calculations
to future work.



\section{Dark Matter Production Mechanisms}
\label{sec:dmprod}

Light bosonic DM can be produced through a range of early-universe
 mechanisms, the simplest of which is purely gravitational production
during inflation. For a spin-0 field, inflation `stretches' quantum
fluctuations to super-horizon scales. After many $e$-folds of inflation, this
results in the observable universe having, at the end, approximately the same
background value of the field everywhere.
This background value persists until the universe cools down enough
that the Hubble rate is less than the mass of the particle, after
which the scalar field starts oscillating, and behaves like matter.

In many models, axions have a periodic potential.
For a field with a cosine potential, $V(a) = m_a^2 f_a^2 \cos(a/f_a)$,
and a mass that does not change with temperature, its
DM abundance today is~\cite{Irastorza:2018dyq}
\begin{equation}
\frac{\Omega_{a}}{ \Omega_{\rm DM}} \simeq \left(\frac{m_a}{\eV}\right)^{1/2}\left(\frac{f_a}{1.5\times 10^{11}\GeV}\right)^{2}\,\left(\frac{\theta_i}{\pi/\sqrt{3}}\right)^2,
\end{equation}
with $\theta_i \equiv a_i / f_a$, where $a_i$ is the post-inflationary
value of the field over the observable universe. 
The scale $f_a$ is generally associated with the same physics
that result in axion-SM couplings, which are then suppressed by $\sim 1/f_a$.
Hence,
the mass and couplings ranges that our experiments will probe
can naturally result in the correct DM abundance, through
this `misalignment mechanism'.

The QCD axion is slightly different from this case, both because its mass
and symmetry-breaking scale are related, and also because its potential
is temperature-dependent \cite{diCortona:2015ldu}. Consequently, misalignment production
can only account for all of the DM if $m_a \lesssim 0.6 \meV$, well 
below the mass range we have been considering. However, there are
post-inflationary production mechanisms that can increase this abundance.
In particular, if the axion only becomes an effective degree of freedom 
post-inflation, then the phase transition in which this occurs
will generically create a network of strings, whose cores
contain the unbroken phase. This string network will
evolve until around the QCD phase transition, when the temperature-dependent
axion mass becomes comparable to the Hubble rate. At this point,
it is expected to decay through the formation of domain walls.

Since the string network is expected to evolve towards an attractor solution,
with eventual statistical properties almost independent of the
post-phase-transition configuration,
the axion DM abundance should theoretically 
depend only on the QCD axion mass.\footnote{This is
true if the domain wall number is 1; if it is larger, than domain
walls are long-lived, which can cause cosmological problems.} However, each of the stages
involves complicated, non-equilibrium physics, and plausible predictions
for the end abundance can differ by several orders of magnitude~\cite{ed,Davis:1986xc,Hagmann:2000ja,Hiramatsu:2010yu,Klaer:2017ond}.
In theories with extra forms of new physics, there can also be
other production mechanisms which give a full DM
abundance of QCD axions at high masses, including
the range we have considered~\cite{Kawasaki:2014sqa,Ringwald:2015dsf,Co:2017mop}.
These cosmological uncertainties motivate searching
for QCD axion DM across as wide a mass range as possible.

Spin-1 DM can also be produced during inflation, through a process
similar to the spin-0 misalignment mechanism. Once again, quantum
fluctuations are blown up by inflation to super-horizon scales.
These do not start oscillating until the Hubble rate becomes smaller
than the DM mass. The difference from the scalar case is that,
during the time when the field is not oscillating, the 
magnitude of the vector field still decreases as $a^{-1}$,
where $a$ is the FRW scale factor, and so the energy density in the field
decreases as $a^{-2}$. In contrast, the potential energy density in a scalar
field does not change until it starts oscillating.
Consequently, the vector modes which are redshifted least are
those which enter the horizon just as they are becoming non-relativistic.
Larger-scale modes spent more time being redshifted before starting to oscillate, while smaller-scale modes spend time red-shifting rapidly
as radiation. At late times, this results in a DM population that is dominated
by modes at that special comoving scale, giving an overall
density of~\cite{Graham:2015rva}
\begin{equation}
\frac{\Omega_{\rm DP}}{ \Omega_{\rm DM}} \sim \left(\frac{m_A}{\eV}\right)^{1/2}\,\left(\frac{H_I}{5\times 10^{12}\GeV}\right)^2,
\end{equation}
where $H_I$ is the Hubble scale during inflation. 
Bounds on the CMB tensor to scalar ratio constrain $H_I \lesssim 10^{14} \GeV$
\cite{Ade:2015lrj}, so this production mechanism can produce all of the DM
at masses $\gtrsim 10^{-5} \eV$.


\section{Discussion}
\label{sec:discussion}

In this work, we have outlined an experimental proposal for light
bosonic dark matter searches in the $\sim \eV$ mass range.
Using periodic, layered dielectrics along with modern
low-dark-count photodetectors, these could have sensitivity
to significant areas of new parameter space for dark photon
and axion dark matter.
Advantages of our proposal include simple target materials, small
target volumes, and signal power emitted into collimated IR-UV photons, which
can be efficiently detected.

The closest existing experimental proposals are the `dish antenna'
searches~\cite{Horns:2012jf,Suzuki:2015sza}, which use a mirror to convert
DM to collimated photons. Modeling a mirror as a very-high-refractive-index
material, these are effectively a one-interface version of a dielectric
haloscope, and consequently, only efficiently `use' a single-wavelength-high
volume above the mirror.
In contrast, dielectric haloscopes can achieve near-optimal dark matter
absorption rates per unit volume (and, for axions, at a given background
$B$ field), averaged over a range of possible dark matter frequencies.
Achieving these high rates relies on
constructing robust dielectric layers with high refractive index
contrasts and large numbers of layers. 
At optical frequencies, it is relatively simple to operate in the
regime where (thermal) backgrounds are small, so our fairly low
$Q$ factors allow for broad mass coverage without sacrificing
too much signal-to-noise.

The search for new light particles is ongoing on
many fronts, including non-dark matter probes of new
particles around the $\eV$ scale. Improved solar
observations ~\cite{Raffelt:1996wa} and dark photon
helioscopes could probe some of the
same parameter space. The proposed IAXO axion helioscope
experiment~\cite{Armengaud:2014gea} and the ALPS II light
shining through walls experiment \cite{Bahre:2013ywa}
would improve on current axion-photon coupling bounds
at masses $\lesssim 0.3 \eV$ and $\lesssim \mbox{few} \times 10^{-4} \eV$,
respectively. For axion couplings to fermions, the
ARIADNE~\cite{Arvanitaki:2014dfa} experiment aims to search
for new spin-dependent forces, and could
probe axion-nucleon couplings for masses $\lesssim 10^{-2}
\eV$. A signal found in a DM experiment would strongly
motivate searches of these kinds, while conversely, signals
in non-DM experiments could inform the dark matter search
program.

The frequency coverage of our proposal is complementary to existing
dark matter direct detection searches: at DM masses above
$\sim 10 \eV$, absorptions have enough energy to ionize atoms in
convenient target materials such as liquid xenon. Such detectors
have the advantage of large target volumes, and easy detection of
ionizations, enabling them to place strong constraints on dark photon
DM~\cite{An:2014twa,Bloch:2016sjj,Agnese:2015nto,Aprile:2016wwo,Hochberg:2016sqx}.
Direct detection experiments with lower thresholds are becoming
possible, extending future reach toward the $\eV$ scale
\cite{deMelloNeto:2015mca,Tiffenberg:2017aac}.
At DM masses $\lesssim \eV$, a number of other types of collective low-lying
excitations have been proposed for DM absorption
\cite{Hochberg:2016sqx,Hochberg:2016ajh,Hochberg:2017wce,Arvanitaki:2017nhi,Knapen:2017ekk,Bunting:2017net,Horns:2012jf}.

As reviewed in the previous section, the isotropic
target materials we have considered are generally not optimal
for detecting DM particles with different couplings. 
In recent work, \cite{Arvanitaki:2017nhi} put forward
a DM-to-photon conversion scheme based on a gas-phase target, which
could probe a range of DM candidates and interactions through molecular
excitations in the $\sim 0.2-20\eV$ range.


A clear extension of our proposal is toward lower frequencies. The
ultimate aim would be to close the sensitivity gap in axion-photon
couplings between microwave proposals such as MADMAX, and the
optical-frequency experiments we have described.
In particular, the best-motivated candidate for light bosonic
dark matter is the QCD axion; the mass range which can give the
full dark matter abundance is very broad given large theoretical
uncertainties and different cosmological histories. Astrophysical
constraints imply that the Peccei-Quinn scale for a generic QCD axion is $\gtrsim 10^8
\GeV$~\cite{Chang:2018rso}, corresponding to masses $\lesssim 60
\meV$.
Similarly, dark photon DM at lower frequencies is currently
poorly constrained, and as discussed in Section~\ref{sec:dmprod}, could
be produced during high-scale inflation.

Photon detection at these mid- to far-IR frequencies is more difficult
than in the optical, but even with bolometric detectors, photonic
haloscopes could reach interesting new parts of parameter
space. For example, axion dark matter with the KSVZ photon coupling
would produce a signal power of $\sim 10^{-22} \, {\rm W}$
(independent of $m$)
from a half-wave stack of 1000 high-contrast layers, with area $(10 \cm)^2$, in a 10 Tesla $B$ field. The same layers, in the absence of a $B$ field,
would produce a signal power of $\sim 4 \times 10^{-12} \, {\rm W}$ at meV
energies, 
from dark photon DM with $\kappa = 10^{-10}$ (which is the approximate
astrophysical bound in the $\sim \meV$ range).
For comparison, bolometers with noise equivalent powers of $\sim
10^{-20} {\rm W} / \sqrt{\Hz}$ are achievable with current technology~\cite{Farrah:2017moj,Bradford2015}.
Blackbody radiation backgrounds become more important at these frequencies.
 If the layers are cooled to $T \ll m$, then most of the blackbody 
power is below the signal frequency and could be filtered out before
reaching the detector. For example, only $\sim 10^{-12}$ of the emitted
blackbody power from a stack at $4 \kelvin$ is at frequencies $> 10
\meV$; for the parameters from above, this corresponds to $\sim 3 \times
10^{-25} \watt$. At smaller $m$, better than $\OO(1)$ fractional
frequency selectivity would be necessary. Other experimental challenges
include materials, optics, layer construction, and background rejection.
We leave detailed consideration of experiments at these frequencies to
future work.


\begin{acknowledgments}

We thank Nick Agladze, Francesco Arneodo, Karl Berggren, Aaron Chou, Adriano Di Giovanni, Savas Dimopoulos, Shanhui Fan, Andrew Geraci, Giorgio Gratta, Edward Hardy, Onur Hosten, Kent Irwin, Friederike Janusch, Axel Lindner, Ben Mazin, David Moore, Sae Woo Nam, Maxim Pospelov, Leslie Rosenberg, Gray Rybka, and Mark Sherwin for helpful discussions. We are grateful to Asimina Arvanitaki, Saptarshi Chaudhuri,  Ken Van Tilburg and Xu Yi for helpful discussions and comments on the manuscript, and to Asimina Arvanitaki, Matthew Kleban, and Ken Van Tilburg for introductions to experimental colleagues and collaborators. We gratefully acknowledge the Kavli Institute for Theoretical Physics for hospitality during part of the project.  This research was supported in part by the National Science Foundation under Grant No. NSF PHY-1748958. MB is supported in part by the Banting Postdoctoral Fellowship. Research at Perimeter Institute is supported by the Government of Canada through Industry Canada and by the Province of Ontario through the Ministry of Economic Development \& Innovation.
\end{acknowledgments}


\bibliography{layers}

\end{document}